\documentclass[review]{elsarticle}
\journal{Information and Software Technology}

\usepackage{paralist}
\usepackage{booktabs}
\usepackage[linesnumbered,noend, noline]{styles/algorithm2e}
\usepackage{tikz}
\usepackage{styles/pgf-pie}
\usetikzlibrary{positioning,shadows}
\newif\ifpienumberinlegend
\pgfkeys{/number in legend/.code=
    \expandafter\let\expandafter\ifpienumberinlegend
    \csname if#1\endcsname
    \ifpienumberinlegend

    \def\beforenumber##1\afternumber{}%
    \fi,
    /number in legend/.default=true
}

\definecolor{1c1}{RGB}{188,162,6}
\definecolor{1c2}{RGB}{137,129,80}
\definecolor{1c3}{RGB}{239,167,31}
\definecolor{1c4}{RGB}{88,194,241}
\definecolor{1c5}{RGB}{6,180,188}

\tikzset{mynode/.style={draw=white,solid,circle,fill=green,inner sep=1pt, thick,
text=black}}
\tikzset{arrow line/.style={dashed, line width= 2.5pt, color=#1}}
\usepackage{subfig}

\usepackage{algpseudocode}
\usepackage{url}

\newlength{\bibitemsep}
\newlength{\bibparskip}
\let\oldthebibliography\thebibliography
\renewcommand\thebibliography[1]{%
  \oldthebibliography{#1}%
  \setlength{\parskip}{\bibitemsep}%
  \setlength{\itemsep}{\bibparskip}%
}

\makeatletter
\g@addto@macro{\UrlBreaks}{\UrlOrds}
\makeatother
\usepackage{hyperref}
\usepackage{breakurl}
\newcommand{\urls}[1]{{\scriptsize\url{#1}}}

\def\it{\textit}

\usepackage{wrapfig}

\def\bf{\textbf}

\def\fig {Figure~}

\newcommand{\nd}{\vspace{1mm}\noindent}

\usepackage{balance}  
\usepackage{graphics} 
\usepackage{booktabs}
\usepackage{ccicons}

\usepackage{url}
\usepackage{fancybox}
\usepackage{multirow}
\usepackage{flushend}
\usepackage{booktabs}
\usepackage{tabularx}
\usepackage{comment}
\usepackage{array}
\usepackage[flushleft]{threeparttable}
\usepackage{mdframed}
\graphicspath{{}{images/}{dia/}}
\DeclareGraphicsExtensions{.pdf,.png}

\usepackage{listings}

\newcounter{o}
\newcounter{d}
\newcounter{t}
\usepackage{tikz}
\definecolor{1c1}{RGB}{188,162,6}
\definecolor{1c2}{RGB}{137,129,80}
\definecolor{1c3}{RGB}{239,167,31}
\definecolor{1c4}{RGB}{88,194,241}
\definecolor{1c5}{RGB}{6,180,188}

\tikzset{mynode/.style={draw=white,solid,circle,fill=green,inner sep=1pt, thick,
text=black}}
\tikzset{arrow line/.style={dashed, line width= 2.5pt, color=#1}}

\def\bf{\textbf}

\def\fig {Figure~}

\def\tbl {Table~}

\def\sec {Section~}

\def\it{\textit}

\usepackage{paralist}
\usepackage{hyperref}
\usepackage{soul}

\normalsize

\usepackage{enumitem}

\usepackage{caption}
\usepackage{tikz}

\usepackage{pdflscape}
\lstdefinestyle{inlinecode}{basicstyle={\ttfamily\scriptsize\bfseries}}

\usepackage{tcolorbox}

\usepackage{paralist}
\usepackage[outercaption]{sidecap}    
\usepackage[figuresright]{rotating}

\usepackage{footmisc}

\usepackage{enumitem}
\newcolumntype{C}[1]{>{\centering\let\newline\\\arraybackslash\hspace{0pt}}m{#1}}
\newtcolorbox
{mybox}[2][]{colbacktitle=red!10!white,
colback=blue!10!white,coltitle=black!70!black,
title={#2},fonttitle=\bfseries,#1}

\AtBeginDocument{%
  \providecommand\BibTeX{{%
    \normalfont B\kern-0.5em{\scshape i\kern-0.25em b}\kern-0.8em\TeX}}}
    
\title{Early Prediction for Merged vs Abandoned Code Changes in Modern Code Reviews}

\begin{document}

\begin{frontmatter}

\author{Khairul Islam$^a$, Toufique Ahmed$^b$, Rifat Shahriyar$^a$, Anindya Iqbal$^a$, and Gias Uddin$^c$}
\address{$^a$Bangladesh University of Engineering and Technology, $^b$University of California, Davis and $^c$University of Calgary}

\begin{abstract}
\noindent\textbf{Context:} 
The modern code review process is an integral part of the current software development practice. Considerable effort is given here to inspect code changes, find defects, suggest
an improvement, and address the suggestions of the reviewers. In a code review process, several iterations usually take place where an author submits code changes and a reviewer gives feedback until is happy to accept the change. In around 12\% cases, the changes are abandoned, eventually wasting all the efforts.

\noindent\textbf{Objective:}  
In this research, our objective is to design a tool that can predict whether a code change would be merged or abandoned at an early stage to reduce the waste of efforts of all stakeholders (e.g., program author, reviewer, project management, etc.) involved.  The real-world demand for such a tool was formally identified by a study by Fan
et al. \cite{fan2018early}.

\noindent\textbf{Method:}  
We have mined 146,612 code changes from the code reviews of three large and popular open-source software and trained and tested a suite of supervised machine learning classifiers, both shallow and deep learning-based. We consider a total of 25 features in each code change during the training and testing of the models. The features are divided into five dimensions: reviewer, author, project, text, and code.

\noindent\textbf{Results:}
The best performing model named PredCR (Predicting Code Review), a LightGBM-based classifier achieves around 85\% AUC score on average and relatively improves the state-of-the-art \cite{fan2018early} by 14-23\%. In our extensive empirical study involving PredCR on the 146,612 code changes from the three software projects, we find that \begin{inparaenum}[(1)]
\item The new features like reviewer dimensions that are introduced in PredCR are the most informative. \item Compared to the baseline, PredCR is more 
effective towards reducing bias against new developers. 
\item PredCR uses historical data in the code review repository and as such the performance of PredCR improves as a software system evolves with new and more data.
\end{inparaenum} 

\noindent\textbf{Conclusion:} 
PredCR can help save time and effort by helping developers/code reviewers to prioritize the code changes that they are asked to review. 
Project management can use PredCR to determine how code changes can be assigned to the code reviewers (e.g., select code changes that are more likely to be merged for review before the changes that might be abandoned).

\end{abstract}

\begin{keyword}
Code Review\sep Patch\sep Early Prediction\sep Merged \sep Abandoned
\end{keyword}

\end{frontmatter}

\section{Introduction}\label{sec:introduction}

\begin{figure}[htbp]
\centering
  \includegraphics[scale=0.45]{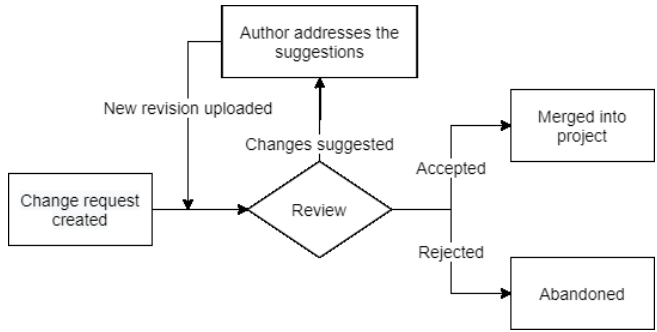}
  \caption{Workflow of a modern code review process}
\label{fig:gerritWorkflow}
	\vspace{-6pt}
\end{figure}

Code review is a practice where a developer submits his/her code to a peer (referred to as `reviewer') to
judge the eligibility of the written code to be included in the main
project code-base. Code review helps remove errors and issues at the early
stage of development. As such, code review can reduce bugs very early and improve software quality in a cost-effective way.
A code review process has some distinct steps (see \fig\ref{fig:gerritWorkflow}). The process starts when a
developer introduces a code change by creating a patch or revision. The
developer or the project moderator assigns a reviewer to examine this change
request~\cite{xia2015elblocker}. The reviewer inspects the code, discusses any possible
improvement, and often suggests fixes. After the review, the developer may
provide a new patch or revision addressing the review comments and generate a
new review iteration. This process repeats until either the reviewer accepts the
changes and it gets merged to the project, or the reviewer rejects the code changes and it gets abandoned~\cite{jeong2009improving}. Such a workflow is facilitated by different automated code review tools such as 
Gerrit \cite{bacchelli2013expectations}. 

Substantial efforts are spent by code reviewers to review a patch
thoroughly, to make code changes, and to analyze comments/suggestions made by
the authors. If a change is abandoned after some iterations, it causes
significant waste of time and resources both for the code reviewer and the code
author. Indeed, we found that around 12\% of code changes are abandoned in our
mined data in three large and popular open-source software projects (see \tbl\ref{table:dataCollection} in \sec\ref{sec:empiricalStudy}). Therefore,
if we can predict early whether a code change would be merged or abandoned in
the long run, we can reduce the waste in effort and time by both code reviewers
and authors.
The prediction has to come as early as possible so that the reviewers can use it
to prioritize which code change to review next. On the other hand, the
management can analyze the cause of an ongoing review process with negative
predictions and intervene to save resources.

The real-world demand for a tool to early predict the future merge/abandon
chance of a code change was previously identified by a study by Fan et al.
\cite{fan2018early}.
They surveyed 59 developers from three popular open-source software communities (Eclipse, LibreOffice, and GerritHub) and asked them whether they needed a tool to predict early if a code change will be merged or abandoned in the future. The developers agreed that they need a
tool to early predict whether a change would be
merged/abandoned in the future. Developers pointed out that this will (1) help
prioritize code changes to review, (2) increase their confidence in merging the
changes, and (3) reduce the resources wasted due to abandoned changes. 

A number of techniques and tools are developed in recent years to assist code reviews
with such early prediction. Jeong et al. \cite{jeong2009improving} proposed a
model to predict patch acceptance in the Bugzilla system. Gousios et al.
\cite{gousios2014exploratory} predicted pull request acceptance on GitHub. They
calculated features when the pull request has been closed or merged.
The most recent early prediction model is a shallow learning Random Forest model
developed by Fan et al. \cite{fan2018early}. They compared their performance
with Jeong et al.~\cite{jeong2009improving} and Gousios et
al.~\cite{gousios2014exploratory} and showed better results at predicting merged
changes. 

Unfortunately, all the above approaches suffer from one or more of the following 
shortcomings: \begin{inparaenum}[(1)]
\item  Jeong et al. \cite{jeong2009improving} used programming language-specific keywords as features but did not use any historical
data, which can offer more contexts to predict the likelihood of future merged/abandoned states.
\item Gousios et al. \cite{gousios2014exploratory} predict just before a pull request is merged/abandoned, which might be too late to save efforts because a code review can span over multiple iterations and interactions between the code author and the code reviewer. Intuitively, the sooner we can predict (in this cycle of iterations), the more efforts and time we can hope to save for both stakeholders.
\item Fan et al. \cite{fan2018early} do not use any reviewer or project-related dimensions, which can offer useful insights that are more specific to a reviewer or a project.
\item All the three models also suffer from bias against new authors, i.e., pull requests from new authors could be unfairly predicted as most likely to be abandoned due to 
lack of data.
\end{inparaenum} Therefore, software developers and code reviewers can benefit from a more robust
tool that can more reliably predict whether a code change would be
merged or abandoned in the future.

In this paper, we have conducted an empirical study on the feasibility of
developing a better classification model by addressing the above limitations of
prior works. Using Gerrit API, we have
mined 146,612 code changes from the code reviews of three large and popular
open-source software projects (Eclipse, LibreOffice, and GerritHub). Each code change has information of whether it is merged or
abandoned - this is our target variable. For
each code change, we compute 25 features from five dimensions: reviewer, author, project,
text, and code. We then train and
test five shallow machine learning models and one deep neural network model on the dataset. 
We find that a LightGBM-based model offers the best overall 
performance. We name the model PredCR. In an empirical study involving PredCR, we answer the following five research questions:

\nd\bf{RQ1. Can our proposed early prediction model PredCR outperform the state-of-the-art baselines?}
This validates the contributions of our work, compared to the prior works (\sec\ref{sec:crossValidation}). 
The most recent model on the early prediction of
merged code changes was developed by Fan et al.
\cite{fan2018early}. Their model outperforms previous works (Jeong et al.
\cite{jeong2009improving}, Gousios et al. \cite{gousios2014exploratory}, etc.). 
We have compared our model performance with the state-of-the-art by reproducing their work. We found that PredCR relatively improves the AUC score by 14-23\%. 
The normalized improvements~\cite{fan2018early,costa2016tipmerge} are 44 - 54\%. Therefore, our developed early prediction model PredCR 
offers considerable performance improvement over state-of-the-art baseline (i.e., Fan et al.
\cite{fan2018early}).

\nd\bf{RQ2. How effective is each feature dimension in our proposed approach?}
This investigates how each feature dimension in PredCR performs. We
have found that (Section \ref{sec:featureDimension}) on average the AUC scores
in models on the reviewer, author, project, text, and code dimensions are 77\%, 67\%, 58\%, 53\%, and 57\% respectively. So previous experience-related features have much impact on the code change outcome. Also, when dimensions are used all together the average AUC score is around 85\%. This validates that PredCR benefits from using all the dimensions.

\nd\bf{RQ3. How well does the model handle bias against new authors?}
As we noted before, state-of-the-art tools to predict early merge/abandoned changes suffer from 
bias against new authors. One of our goals while designing PredCR was to reduce such bias so that we can facilitate better onboarding of new reviewers and authors into the software ecosystem. 
We have used historical data to predict merged code changes whereas new authors
have little prior records in the system. We find (Section \ref{sec:newAuthor}) that
PredCR achieves on average 78.7\% AUC score for new authors. This relatively improves the AUC scores by 21-30\%  compared to the state-of-the-art  \cite{fan2018early}.

\nd\bf{RQ4. How well does our approach work while using multiple revisions?}
In real life, code changes generally go through multiple revisions before
finally getting merged or abandoned. Intuitively, it is more difficult to predict the merge/abandoned change of a code change if we are only looking at the first revision, compared to the last revision. As such, it would be beneficial to find whether and how PredCR can improve its prediction accuracy as we add more revisions to it over time. This research question leads to exploring
how well PredCR performs when predictions are updated at each new revision of
the same code change. We find (Section \ref{sec:multipleRevision}) that if features related to
prior revisions are added to the feature set, 6-15\% relative improvements are achieved in
terms of the AUC score in the last revision compared to the first. Therefore, PredCR achieves better performance during the latter stages of a revision chain. 

\nd\bf{RQ5. How well does the model improve over time?}
As a software project evolve, it can have more data to train over time. Therefore, it is important to
understand whether PredCR is able to improve its prediction accuracy, as a project evolves. We thus sliced a project data by time into 11 folds, where fold 0 contains the earliest data and fold 10 contains the most recent data. We
find (Section \ref{sec:improvementOverTime})  that PredCR gives 5-9\% better
AUC scores in the second half of the folds (i.e., folds 5-10) than the first half of the folds (i.e., folds 0-4). Therefore, the performance of PredCR for an evolving software project improves over time, as we have access to more data of the software.

Our tool can help reviewers manage their review works better. It can also assist
project management to make decisions regarding resource allocation. Code changes
with the possibility of being merged into the main codebase can be given more
focus than those predicted as abandoned. PredCR also extracts features to understand change intent: bug fix, feature implementation or refactoring (Section \ref{textFeature}). In practice, bug fix changes have more importance than feature implementations and feature implementations have more importance than refactoring. So reviewers can use PredCR to label more important code changes (e.g., bug fixes). Then prioritize changes that are more important and have better merge probability.
Table \ref{table:contributions} summarizes the contributions we have made in this
work. The usage scenarios of our proposed tool are as below:

\begin{table}[!htb]
\centering
\caption{Research contributions made in this work}
\label{table:contributions}
\resizebox{\textwidth}{!}{
\begin{tabular}{p{2cm}|p{5cm}|p{5cm}}\toprule

\textbf{Topic} & \textbf{Research Contribution} & \textbf{Research Advancement}  \\ 
\midrule
Prioritizing review requests & Our work shows considerable performance improvement compared to the state-of-the-art \cite{fan2018early} in early predicting outcome of code
changes. & Predicting outcome of code changes has
been highlighted by many prior studies \cite{jeong2009improving}
\cite{jiang2013personalized} \cite{gousios2014exploratory}
\cite{thongtanunam2017review} \cite{fan2018early} \cite{zhao2019improving}. Our
study will help to reduce the difficulties programmers are facing in the rapid
growth of software projects.  \\ 
\midrule Reducing prediction bias & We
have shown that PredCR can reduce the prediction bias against new authors is
most cases compared to the state-of-the-art \cite{fan2018early}. & Code review
approach for newcomers is different \cite{kovalenko2018code}. So careful
approach is necessary so that such a prediction model does not discourage them
from contributing. Our study will help the community by reducing this bias. \\ 
\midrule Update prediction at multiple revisions & We
have presented an adjusted approach that can update prediction at the submission
of new revision for a code change so that efforts at later revisions are
recognized. & Compared to prior arts which calculates code change related
features only at initial submission \cite{fan2018early} or just before closing
\cite{jiang2013will} \cite{gousios2014exploratory} \cite{thongtanunam2017review},
our approach adds the flexibility to also consider subsequent revisions. This is more useful as it scores based on the latest patch before any review has started on it. \\ 
\bottomrule
\end{tabular} 
}
\end{table}

\begin{itemize}[leftmargin=10pt]
\item\textbf{Without PredCR:} Bob is a developer in a large project team. His
responsibility is to review submitted code changes by other developers. With the
expansion of the projects, the number of code changes he has to review has increased
too. He inspects the code changes serially by the order of submission time or
randomly. However, it is difficult for him to keep the focus on reviewing so many code
changes. Also, code changes with better quality are often taking much longer to
merge into the project for falling behind in the queue. Some of the code changes
are being abandoned even after his effort and time. Also after giving some
initial reviews in a code change, he has to go through it again to check if the
author has improved it in a later revision.

\item\textbf{With PredCR:} Bob and his team adopt our tool. The tool predicts the
merged probability of code changes that are assigned to him. Now he can prioritize the code changes based on their probability of getting merged. So he can focus more on
those with a better chance of getting merged in the future. He can also use the tool features to filter more important changes (e.g., bug fixes) and prioritize only them. Also, there would be less delay for the
better code changes, as they will be reviewed and accepted earlier. As such, Bob now can spend less time on code changes that will likely be abandoned in the long run. Moreover, the tool updates the
prediction with each new revision/patch submission of the same code change. This helps Bob refine his decision to prioritize code changes for review, e.g., a code author may radically improve a new version of a code that was previously predicted to be abandoned by PredCR. With new data, PredCR can update its prediction that the updated code has now more chance of getting merged than abandoned. This will help Bob to then focus more effort on the new code changes during reviews. 
\end{itemize}


\nd\bf{Replication Package.} \urls{https://github.com/khairulislam/Predict-Code-Changes}.

\nd\bf{Paper Organizations.} The rest of the paper is organized as follows.
Section \ref{sec:related-work} presents the prior works related to ours. Section
\ref{sec:empiricalStudy} presents the data collection process, studied
features, research questions, and evaluation metrics. Section \ref{sec:result}
presents the answers to the research questions presented in the previous
section. Section \ref{sec:discussions} discusses the major themes of our study
results and highlights the finding of our study.
Then in Section \ref{sec:threatsToValidity}, we have presented the threats to the validity of our work. And Section \ref{sec:conclusion} has the concluding
remarks.

\section{Related Work}\label{sec:related-work}
In this section, we have presented the prior works related to our study. We have discussed their motivations, working setups, features used, and limitations. Table \ref{table:relatedWorks} shows the summary of those works and our comparison with them.

\begin{table*}[!htb]
\centering
\caption{Comparison of our paper with related works}
\label{table:relatedWorks}
\resizebox{\textwidth}{!}{
 \begin{tabular}{p{1.6cm}|p{4cm}|p{4.2cm}|p{4.2cm}}\toprule

\textbf{Topic} & \textbf{Our works} & \textbf{Prior study} & \textbf{Comparison} \\ 
\midrule
Early Prediction in Code Reviews & Our goal is to predict early whether a code change will be
merged or abandoned to prioritize reviews and to reduce waste of efforts on
abandoned changes. &  Predict whether a patch will be accepted
\cite{jeong2009improving, gousios2014exploratory, hellendoorn2015will}, will need more than one submission to be accepted
\cite{huang2019would, gerede2018will}, will fail to attract reviewers
\cite{thongtanunam2017review}, will be closed earlier than others
\cite{zhao2019improving}. Early prediction of a code change being
merged \cite{fan2018early}.  & Our goal is to predict merge probability early
before any review starts. Similar to Fan et al. \cite{fan2018early}. \\ 
\midrule
Review tool used & Gerrit code review tool & Bugzilla \cite{jeong2009improving},
Linux kernel \cite{jiang2013will}, Github \cite{gousios2014exploratory, hellendoorn2015will, zhao2019improving},  Gerrit 
\cite{fan2018early, gerede2018will, huang2019would} & As all features
available on one tool, might not be available on another, our work on Gerrit can
not be compared directly with all of them. \\ 
\midrule
Feature dimensions used & Reviewer, author, project, text and code related
features. Experience related features were calculated using more recent data
(past 60 days). & Code or patch \cite{jeong2009improving}\cite{jiang2013will}
\cite{huang2019would} \cite{gousios2014exploratory} \cite{fan2018early}, bug
report \cite{jeong2009improving}, project \cite{gousios2014exploratory, thongtanunam2017review, fan2018early}, author
\cite{gousios2014exploratory} \cite{fan2018early}, review \cite{jiang2013will}
\cite{thongtanunam2017review}, text \cite{thongtanunam2017review}
\cite{fan2018early}, reviewer \cite{thongtanunam2017review}  & We have focused
on more recent performance of authors and reviewers. All features presented by
us in Section \ref{sec:studiedFeatures} are available from the creation of the
code change. \\
\midrule
Program language dependency & We have not used any language-dependent
features & Jeong et al. \cite{jeong2009improving}, Hellendoorn et al.
\cite{hellendoorn2015will}, and Huang et al. \cite{huang2019would} used features
dependent on Java language. & PredCR can be used on any project using the Gerrit
tool as it is language-independent.  \\ 
\bottomrule
\end{tabular}
}
\end{table*}

\subsection{Early Prediction in Code Reviews}
Jeong et al. \cite{jeong2009improving} focused on predicting patch acceptance at any state of revisions. They suggested that patches predicted as accepted can be auto-accepted and authors can use it before submitting a patch to get feedback on it. Also, reviewers can use it to predict patch quality. Jiang et al. \cite{jiang2013will} conducted a study on the Linux kernel and examined the relationship between patch characteristics and patch reviewing/integration time. Kamei et al. \cite{kamei2013large} built a change risk model based on characteristics of a software change to predict whether or not the change will lead to a defect. However, this doesn't predict whether the change will be eventually merged or abandoned. Gousios et al. \cite{gousios2014exploratory} predicted acceptance of pull requests. To obtain an understanding of pull request usage and to analyze the factors that affect such development. 

Hellendoorn et al. \cite{hellendoorn2015will} used natural language processing techniques to compute how similar a code change is to previous ones. They then predicted whether it will be approved based on the review outcomes of similar ones. Thongtanunam et al. \cite{thongtanunam2017review}  investigated the characteristics of patches that: (i) do not attract reviewers, (ii) are not discussed, and (iii) receive slow initial feedback. They calculated features just before the code change was closed and predicted acceptance for it at that moment. Gerede et al. \cite{gerede2018will} focused on predicting whether or not a code change would be subject to a revision request by any of its reviewers. 

Fan et al. \cite{fan2018early} predicted whether a code change will be merged or abandoned as soon as it was submitted. Their main objective was to prioritize the code review process by early predicting code changes that are more likely to be merged. They compared their works with Jeong et al. \cite{jeong2009improving}, Gousios et al. \cite{gousios2014exploratory}, and show state-of-the-art performance. Zhao et al. \cite{zhao2019improving} proposed a learning-to-rank (LtR) approach to recommending pull requests that can be quickly reviewed by reviewers. Different from a binary model for predicting the decisions of pull requests, their ranking approach complements the existing list of pull requests based on their likelihood of being quickly merged or rejected.  Huang et al. \cite{huang2019would} proposed a method to predict the time-cost in code review before a submission is accepted. They focused on predicting whether a submission will be accepted on the first submission and whether it will take more than 10 submissions.  

Our target is to predict the merged probability of a code change request as soon as it is submitted before any review has come. This is similar to the work by Fan et al. \cite{fan2018early}.

\subsection{Review tool used}
Jeong et al. \cite{jeong2009improving} used the Bugzilla system in Firefox and the Mozilla Core projects. Gousios et al. \cite{gousios2014exploratory}, Zhao et al. \cite{zhao2019improving}, Hellendoorn et al. \cite{hellendoorn2015will}  worked with pull requests in GitHub projects. Jiang et al. \cite{jiang2013will} worked on the Linux kernel which is supported by Git repositories.  Thongtanunam et al. \cite{thongtanunam2017review},Huang et al. \cite{huang2019would}, Gerede et al. \cite{gerede2018will}, and Fan et al. \cite{fan2018early} worked on open source projects using the Gerrit tool. We have also worked with the Gerrit tool. 

\subsection{Feature dimensions}
Jeong et al. \cite{jeong2009improving} used patch metadata, patch content, and bug report related features. Bug report-related features are very specific to the Bugzilla system they worked on. However, they do not use any historical data in the feature set. Shin et al. \cite{shin2009does} showed that without historical data fault prediction models usually have low performance. Gousios et al. \cite{gousios2014exploratory} used pull request, project, and developers' characteristics-related features. Both Jeong et al.\cite{jeong2009improving} and Gousios et al. \cite{gousios2014exploratory} used some features (time after open) which are not available when the first patch is submitted. Gousios et al. \cite{gousios2014exploratory} also used review activities in previous revisions in the feature set (num\_comments, num\_participants).  They calculated features at the time a pull request has been
closed or merged.

Jiang et al. \cite{jiang2013will} grouped the features into six dimensions: experience, email, review, patch, commit, and development. The review group is related to review participation in the prior patches. The email feature contains information related to prior patches. Thus many of the features are not available when submitting the first patch. Kamei et al. \cite{kamei2013large}
grouped the features into diffusion, size, purpose, history, experience dimensions. Thongtanunam et al. \cite{thongtanunam2017review} extracted patch metrics in five dimensions: patch properties, history, past involvement of an author, past involvement of reviewers, and review environment. Their history feature is related to review activities in prior patches of the patch set. They calculated the features just before the code change was merged or abandoned. Fan et al. \cite{fan2018early} grouped the features into five dimensions: code, file history, owner experience, collaboration network, and text. All of these features are available when the first revision of the code change request is being submitted. 

We have grouped our features into five dimensions: reviewer, author, project text, code. All of those features are calculated after the first revision is created. 

\subsection{Programming language dependency}
Jeong et al. \cite{jeong2009improving} used Java language-specific keywords in their feature set to predict patch acceptance. Hellendoorn et al. \cite{hellendoorn2015will} trained and tested their language models on pull requests that only contain java files. Huang et al. \cite{huang2019would} used code modifying features and code coupling features which are java language-dependent. Therefore, they filtered out any changes from their dataset which contained any non-java file. These works are programming language-dependent, so can not be used for projects of different languages. Other previous works discussed \cite{fan2018early, thongtanunam2017review, jiang2013will, gousios2014exploratory}, don't have a programming-language dependency.

Our work doesn't use any language-specific features. So it is programming language-independent.

\section{Empirical Study Setup }
\label{sec:empiricalStudy}
In this section, we have described how we have collected the data from Gerrit
projects and preprocessed them before using them in the experiment. Then we have
explained the features extracted from the dataset, which we have grouped into
five dimensions. We have presented the rationale and explained how the
features were calculated. Then, we have described our evaluation metrics to
measure the prediction performance. Finally, we have presented the research questions we shall answer in our work.

\subsection{Data collection and Preprocessing}
\label{dataCollection}
We have used the REST API provided by Gerrit systems to collect data from three
Gerrit projects LibreOffice, Eclipse, and GerritHub. The miner was created following the approach presented by Yang et al. \cite{yang2016mining}. We have collected changes with the status ``merged" or
``abandoned". We have mined a total of 61062, 113427, and 61989 raw code changes respectively from LibreOffice, Eclipse, and GerritHub respectively within the time period mentioned in Table \ref{dataCollection}.

To filter out the inactive/dead sub-projects, we have selected sub-projects with
at least 200 merged code changes. Hence, 4, 64, and  48 sub-projects were left respectively from LibreOffice, Eclipse, and GerritHub.
We have removed code changes where subjects contain the word ``NOT
MERGE" or ``IGNORE" since these will eventually be abandoned. We have also
removed changes where the reviewers are the same as the owners. Some changes didn't have patchset data available anymore, we have also excluded them.  The same
preprocessing steps are applied to all three projects. Table
\ref{table:dataCollection} presents statistics of the finally collected dataset.
We have also collected registration dates for each developer account. It was
later used during feature extraction for the computing experience of the developer. In case of missing values on the date of
registration, we have filled them by linearly interpolating them based on the existing dates and account\_id. For example, if account\_id 3 has registration date missing and the closest previous and next account\_ids are 1 and 5 with registration dates 01-01-2018 and 01-05-2018. Account\_id 3 will be assigned 01-03-2018 as the registration date.

\begin{table*}[htbp] \centering
\caption{Statistics of collected data}
\label{table:dataCollection}
\resizebox{\textwidth}{!}{
\begin{tabular}{llrrr}\toprule
\textbf{Project} & \textbf{Time period} & \textbf{Changes} & \textbf{Merged} & \textbf{Abandoned} \\ 
\midrule
LibreOffice & 2012.03.06 -- 2018.11.29 & 56,241 & 51,410(91\%) & 4,831(9\%) \\ 
Eclipse & 2012.01.01 -- 2016.12.31 & 57,351 & 48,551(85\%) & 8,800(15\%) \\ 
GerritHub & 2016.01.03 -- 2018.11.29 & 33,020 & 29,367(89\%) & 3,653(11\%) \\ 
\midrule
Total & & 146,612 & 129,328(88\%) & 17,284(12\%) \\ 
\bottomrule
\end{tabular}
}
\end{table*}

\subsection{Studied features}\label{sec:studied-features}
\label{sec:studiedFeatures}

We have extracted a total of 25 features from the dataset. 
{All features are calculated when the code change is initially submitted (same as Fan et al. \cite{fan2018early})}. Gousios et al. \cite{gousios2014exploratory}, Jiang et al. \cite{jiang2013will}, Thongtanunam et al. \cite{thongtanunam2017review} calculated all features at the time when a
change has been closed. However, the review process has already been finished by then
and no remedy is effective at that point. Our main goal is to predict the possibility of merging/abandonment for code changes as early as possible. For this reason, we have not used the following dimensions:  history (Thongtanunam et al. \cite{thongtanunam2017review}), review (Jiang et al. \cite{jiang2013will}), commit (Jiang et al. \cite{jiang2013will}). These are not available at the initial stage. Also, in Section \ref{sec:multipleRevision} we have shown that by only adding revision numbers to the feature list, PredCR can give significant performance when the prediction is updated after submission of each new revision.

Some features were not available in the Gerrit system. For example: bug report information (Jeong et
al.\cite{jeong2009improving}), email (Jiang et al. \cite{jiang2013will}). When
calculating past record-related features, we have generally considered recent
performances (in the last 7 or 60 days). Fan et al. \cite{fan2018early} added
'recent' prefix to features that were calculated in the last 120 days. Our approach thus is more restrictive in terms of feature history. 
 Table
\ref{table:studiedFeatures} shows our finally selected feature list and the
rationale behind choosing those. We discuss the features and dimensions below.

\begin{table*}[!htb]
\centering
\caption{List of features. The dimensions which we have used, but were not used by state-of-the-art \cite{fan2018early} are highlighted as bold. The features for which we did not find prior studies using them, are highlighted as bold too.}
\label{table:studiedFeatures}
\resizebox{\textwidth}{!}{
\begin{tabular}{p{1.7cm}|p{7.5cm}|l} 
\toprule
\textbf{Dimension} & \textbf{Rationale} & \textbf{Feature Name}  \\ 
\midrule
\multirow{4}{*}{\textbf{Reviewer}} & \multirow{3}{=}{Reviewers number and their past record affect change outcome \cite{rigby2011understanding} \cite{thongtanunam2017review}.} & \textbf{avg\_reviewer\_experience} \\ 
& & avg\_reviewer\_review\_count \cite{thongtanunam2017review}
\cite{baysal2016investigating}  \\ 
& & num\_of\_reviewers
\cite{thongtanunam2017review} \cite{jiang2013will} \\ 
&& \textbf{num\_of\_bot\_reviewers} \\
\midrule

\multirow{6}{*}{Author} & \multirow{6}{=}{Experienced programmer has low defect
probability
\cite{mockus2000predicting}. Developer's experience significantly impacts on change outcome \cite{jiang2013will} \cite{gousios2014exploratory}. More active developers have a better chance at merging patches \cite{baysal2013influence}.} & author\_merge\_ratio \cite{fan2018early} \\ 
& & \textbf{author\_experience}  \\ 
& & author\_merge\_ratio\_in\_project \cite{fan2018early}  \\ 
& & total\_change\_number \cite{fan2018early} \cite{baysal2016investigating} \\ 
& & author\_review\_number \cite{fan2018early} \cite{jiang2013will}  \\ 
& & author\_changes\_per\_week \cite{thongtanunam2017review} \\ 
\midrule

\multirow{3}{*}{\textbf{Project}} & \multirow{2}{=}{Large workload results in less review participation \cite{baysal2016investigating}.} & project\_changes\_per\_week \cite{thongtanunam2017review} \\ 
& & \textbf{changes\_per\_author} \\ 
& Project's receptiveness affects change outcome \cite{gousios2014exploratory} & \textbf{project\_merge\_ratio} \\ 
\midrule

\multirow{4}{=}{Text} & Well explained descriptions better draw attention\cite{rigby2011understanding} & description\_length \cite{fan2018early}  \\ 
& \multirow{3}{=}{Intent of a code change is related to the kind of feedback
it receives.\cite{wang2019leveraging} } & is\_bug\_fixing \cite{kamei2013large, thongtanunam2017review, fan2018early} \\ 
& & is\_feature \cite{thongtanunam2017review, fan2018early} \\ 
& &  is\_documentation \cite{thongtanunam2017review, fan2018early} \\ 
\midrule

\multirow{8}{*}{Code} & Modifying more directories is usually defect-prone \cite{mockus2000predicting}. & modified\_directories \cite{thongtanunam2017review} \\ 
& Scattered changes are more prone to defects \cite{hassan2009predicting}. &
modify\_entropy \cite{kamei2013large} \\
& \multirow{2}{*}{Larger changes are more defect-prone \cite{moser2008comparative}.} & lines\_added \cite{jeong2009improving} \cite{fan2018early} \\ 
& & lines\_deleted \cite{jeong2009improving} \cite{fan2018early} \\ 

& \multirow{4}{*}{Touching many files is more defect-prone \cite{nagappan2006mining} \cite{moser2008comparative}. }  & files\_modified \cite{gousios2014exploratory}\\ 
& & files\_added \cite{fan2018early} \\ 
& & files\_deleted \cite{fan2018early} \\ 
&& subsystem\_num \cite{fan2018early} \\
 \bottomrule
\end{tabular}
}
\end{table*}

\subsubsection{Feature Dimension 1. Reviewer \label{reviewerFeature}}
Num\_of\_reviewers is the number of human reviewers found in the reviewer list of the
code change. This feature was previously used by Thongtanunam et al.
\cite{thongtanunam2017review} and Jiang et al. \cite{jiang2013will}.
Num\_of\_bot\_reviewers are the number of bot tools added to the reviewer's
list. As these accounts don't actively participate in review discussion but perform different analyses on the patch set, we have kept their number separately. Whether an account is a bot, is determined by checking whether the account name is 'do not use' or it contains any of the following words 'bot’, 'chatbot', 'ci', 'jenkins', or the project name.  We have
calculated a reviewer's experience by the number of years s/he is registered in
this system. We have calculated that using the difference of the revision upload
date and the reviewer's date of registration in this project.  This value is
then averaged by the number of reviewers, which is feature
avg\_reviewer\_experience. A reviewer's review count is found by calculating the
number of closed (merged or abandoned) changes, in the last 60 days, where that
a particular reviewer was involved in the reviewer list. This value is then
averaged by the number of reviewers, which is feature
avg\_reviewer\_review\_count. Thongtanunam et al. \cite{thongtanunam2017review}
introduced similar features that calculated prior patches that a reviewer has
reviewed or authored. 

\subsubsection{Feature Dimension 2. Author \label{authorFeature}}
We have used the recent changes in a 60-day window when calculating author\_merge\_ratio, author\_review\_number, author\_merge\_ratio\_in\_project, 
\\ changes\_per\_week. When calculating the merge ratio, if there are no finished changes of this author, then a default merge ratio of 0.5 is given. Author\_merge\_ratio is the ratio of merged changes among all finished changes created by this author. Author\_review\_number is the number of changes where the author is in the
reviewers' list. Author\_merge\_ratio\_in\_project is the author's merge ratio in the corresponding
sub-project. This sub-project name comes with the "project" key in code change
response, so we have kept it in this way. Changes\_per\_week is the number of
closed changes each week for this author in the last 60 days. Author\_experience is calculated following the same way as the reviewer experience, i.e., taking the difference between the current revision upload date and the author's date of registration in years. Total\_change\_number is the number of changes created by this
author.

\subsubsection{Feature Dimension 3. Project}

We have calculated all project-related features in a 60 days window. \\Project\_changes\_per\_week feature is calculated using the number of changes closed every 7 days among the past 60 days for this sub-project. Changes\_per\_author is the number of closed changes per author in the last 60 days. Project\_merge\_ratio is the ratio of merged and closed changes in the last 60 days for this sub-project. If the project doesn't have any finished changes yet, the default merge ratio of 0.5 is given.

\subsubsection{Feature Dimension 4. Text \label{textFeature}}
These features are calculated on the change description provided for the code
change. The aim is to identify the purpose of the code change. The description is provided in the subject of the code change when it is created. Description\_length is the number of words present in the change description. The other three features have binary values, i.e., 0 or 1. We have marked a code change as documentation if the change description contains "doc", "copyright", or "license". Similarly, we categorize it as bug fixing if the change description contains "bug", "fix" or "defect". Other changes are marked as a feature. These are done following Thongtanunam et al. \cite{thongtanunam2017review}.

\subsubsection{Feature Dimension 5. Code}
This section refers to the features which are related to the changes made in
The source code. Modified\_directories refer to the number of directories modified by this code change. It is calculated by extracting the bottom directories from
file paths. Similarly, subsystem\_num is the number of subsystems (the top directory in the file path) modified in the change. Modify\_entropy is a feature previously proposed by Kamei et al.\cite{kamei2013large}. Entropy is defined as $ -\sum_{k=1}^{n}(p_k * log_2p_k)
$, where $n$ is the number of the files modified and $p_k$ is the proportion of lines modified among total modified lines in this change.
Other features such as files\_added, files\_deleted, files\_modified, and
lines\_added, lines\_deleted are self-explanatory. Most of these source code features have also been used in prior studies \cite{kamei2013large,
gousios2014exploratory, thongtanunam2017review,  fan2018early}.

\subsection{Performance Metrics}
\label{sec:evaluationMetrics}
We use a total of seven metrics to report and compare the performance of PredCR against the baselines in our three datasets. 
The metrics can be broadly divided into two categories: Standard Performance Metrics and Improvement Analysis Metrics.  
All the metrics except one (cost-effectiveness) are used from Python \href{https://scikit-learn.org/stable/index.html}{scikit-learn library}.  The metrics are defined below. 

\subsubsection{Standard Performance Metrics}
We report five standard performance metrics:\begin{inparaenum}[(1)]
\item AUC, 
\item Cost-Effectiveness, 
\item Precision, 
\item Recall, and 
\item F1-score.
\end{inparaenum}  

\nd\bf{AUC.} Area Under the Curve (AUC) of the Receiver Operating Characteristic (ROC) is a
widely used performance measure for prediction models. For our case, the AUC score
calculates the probability that PredCR prioritizes merged code changes more than
abandoned code changes. Following related literature on the early prediction of merged/abandoned code reviews, we use the AUC score to determine the best-performing models. 

\nd\bf{Cost-Effectiveness (ER@K\%).} Cost-effectiveness is used to measure performance given a cost limit. As in
practice, developers can only review a limited number of changes, our target is to correctly predict as many merged cases as possible within that limit. Following prior studies ~ \cite{xia2015elblocker, fan2018early}, we have used EffectivenessRatio@K\% (ER@K\% in short), which evaluates the percentage of merged code changes in the top $K$\% code changes(sorted by decreasing order of merge probability) predicted as "Merged". 

This also helps evaluate how well our model can prioritize the code changes. A larger effectiveness ratio means the model better prioritizes code changes that will eventually be merged. The state-of-the-art \cite{fan2018early} used this metric for the same purpose. Xia et al. \cite{xia2015elblocker} used this metric to evaluate the prioritization of blocking bugs. Jiang et al. \cite{jiang2013personalized} also used this to evaluate the ranking of personalized defect prediction. The authors of these works used prediction probability from the model to prioritize.

By denoting the number of merged changes and the number of changes in top $K$\% as $N_{mk}$ and $N_k$, respectively, we get,

\begin{equation}
ER@K\% = \frac{N_{mk}}{N_k}
\end{equation}

We have used ER@20\% as the default cost-effective metrics. In Section \ref{sec:costEffectiveness}, we have shown PredCR performance when $K$ is varied from 10 to 90. Note that using $K$ at 100 doesn't have any significance, as when choosing top 100\% code changes, the proportion of merged changes and all changes are constant for a test set, irrespective of the model. 

\nd\bf{Precision.} The proportion of changes that are correctly labeled among all predicted examples of that class. For merged and abandoned classes, we presented this metric as P(M) and P(A).
\begin{equation}
P(M) = \frac{TP}{TP+FP},
P(A) = \frac{TN}{TN+FN}
\end{equation}

\nd\bf{Recall.} The proportion of changes that are correctly labeled among changes that actually belong to that class. For merged and abandoned classes, we presented this metric as R(M) and R(A).

\begin{equation}
R(M) = \frac{TP}{TP+FN},
R(A) = \frac{TN}{TN+FP}
\end{equation}

Recall is different from precision in this regard, precision means the percentage of results that are relevant. On the other hand, recall refers to the percentage of total relevant results correctly classified by our algorithm.

\nd\bf{F1-Score.} The harmonic means of precision and recall. For merged and abandoned classes we presented this metric as F1(M) and F1(A).
\begin{equation}
F1(M) = \frac{ 2 * P(M) * R(M)}{P(M) + R(M)}
\end{equation}

\begin{equation}
F1(A) = \frac{2 * P(A) * R(A)}{P(A) + R(A)}
\end{equation}

\subsubsection{Improvement Analysis Metrics}
We report two metrics: \begin{inparaenum}[(1)]
\item Relative Improvement (RIMPR), and  
\item Normalized Improvement (NIMPR).
\end{inparaenum}  

\nd\bf{Relative Improvement (RIMPR).} By relative improvement, we mean the relative change between the two scores. Instead of simply calculating the difference it is better because it considers the difference relative to the old value. For example, improving a score from 20\% to 40\% is only a 20\% increase in score. But only calculating the difference misses the fact that the new score is double the previous score. However, the improvement here is 100\% which clearly shows that fact. Improvement is calculated as follows,

\begin{equation}
    Relative~Improvement~(RIMPR) = \frac{new~ score - old ~score}{old~ score}
\end{equation}
We report this metrics name as RIMPR throughout the rest of the paper.

\nd\bf{Normalized Improvement (NIMPR).} Normalized improvement is a measure proposed by Costa et al. \cite{costa2016tipmerge} 
to evaluate the improvement between two methods in terms of an evaluation
metric. The same metrics have been used by Fan et al. \cite{fan2018early} to
highlight improvements over baselines in prioritizing code changes for
reviewers. It takes room for improvement into consideration. For example:
let us consider accuracy is improved from 80\% to 85\% and F1\_score is improved
from 90\% to 95\%. In both cases, the improvement is 5\%, but normalized
improvement is 25\% and 50\%, respectively. In the latter case, the room for
improvement was only 10\%. Hence, a 5\% improvement here has much more
impact. We have used the short form of this metric as NIMPR.

\begin{equation}
Normalized ~Improvement~(NIMPR) = \frac{new~score - old~score }{1 - old~score}
\end{equation}

\subsection{Experimentation setup and approach \label{sec:experimentationSetup}} 
We have used the longitudinal data setup, previously used by Fan et al. \cite{fan2018early}. Previous works have used similar setups to ensure only using past data to predict future events. Rakha et al.\cite{rakha2017revisiting} used a similar approach in retrieving duplicate issue reports. Bangash et al.\cite{bangash2020time} used time-aware evaluation in cross-project defect prediction.

For each project, the selected code changes are
first sorted in increasing order of creation time.
Then they are divided into 11 non-overlapping windows of equal size. Instead of
traditional ten-fold cross-validation, this approach is followed to ensure that
no future data is used during training.

In the first fold, the model is trained using window 1 and tested on window 2.
In the second fold, the model is trained using windows 1 and 2 and tested on
window 3. Similarly, in the last fold(10), the model is trained on windows 1-10
and tested on window 11. At each stage, we have calculated the AUC, ER@20\%,
precision, recall, and F1 scores for merged and abandoned code changes. Then we
have computed the average of the metrics across ten-folds for both merged and
abandoned code changes. Kaggle kernels were used to run all experiments. They provide an Intel(R) Xeon(R) CPU with 16 Gigabytes of Ram, 4 CPU cores, and a 2.20GHz processor. We have used Python as the programming language. Due to the stochastic nature of the machine learning models, 
it is recommended to run a model multiple times and take the average for final performance 
reporting. In our case, for each model, each experimentation is rerun ten times and the average result is reported to ensure stable model performance. This means that for each model we did longitudinal 10-fold cross-validation 10-times and then took the average. During each of the runs, we did hyperparameter tuning.

\nd\bf{\ul{Model selection process.}} First, we have used StandardScaler to fit and transform the features of each
project. Then to find the best model, we have used six machine learning classifiers
GradientBoosting \cite{friedman2002stochastic}, RandomForest
\cite{breiman2001random}, ExtraTrees \cite{geurts2006extremely},
LogisticRegression, LightGBM \cite{ke2017lightgbm} and Deep Neural Network(DNN). Except for LightGBM and DNN, all
other classifiers are imported from the scikit-learn library.  The
LightGBM classifier used is taken from
lightgbm \footnote{https://lightgbm.readthedocs.io/en/latest/index.html} library. The DNN model was created using the keras\footnote{https://keras.io/} library.

\nd\bf{\ul{Handling class imbalance.}} As this dataset is an imbalanced one, we have considered class imbalance when
training the models. We have balanced the classification loss, by setting the classifier parameter class\_weight to 'balanced'. This uses the values of the target column to automatically adjust weights inversely proportional to class frequencies in the input data. This way class imbalance is taken into consideration when calculating loss. Hence, we have set class\_weight = 'balanced' for all
of these classifiers. Except for GradientBoosting, which automatically
handles class imbalance by constructing successive training sets based on
incorrectly classified examples \cite{friedman2002stochastic}. 

\nd\bf{\ul{Randomness across different runs.}} To introduce randomness across different runs, we set solver = 'saga' for LogisticRegression (suggested by scikit-learn documentation). And subsample=0.9, subsample\_freq=1, random\_state = numpy.random.randint(seed=2021) for LightGBM (this will subsample 90\% of the train data each time). Otherwise, these two models produce the same results after each run, and rerunning them ten times doesn't have a meaning. The DNN model maintains random results because it initializes to random weights. The other models had their random\_state kept to default 'None' during model initialization. We also manually validated whether each run is creating different results.

\begin{figure}[htb]
\centering
  \includegraphics[scale=0.35]{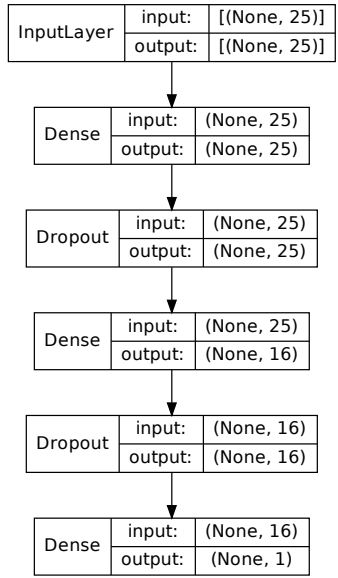}
  \caption{DNN model architecture}
\label{fig:dnn}
	\vspace{-6pt}
\end{figure}

\nd\bf{\ul{Deep Neural Network (DNN) Architecture.}} We have used the deep neural network model to investigate whether neural networks would outperform other machine learning classifiers. The network architecture is shown in figure \ref{fig:dnn}. It contained three dense layers. The input dense layer contains 25 relu units, one for each feature. Then we have added a dropout layer with a 10\% dropout rate, this would randomly drop 10\% of the incoming values, which will help reduce overfit on the training data. Then another dense layer with 16 relu units. Then another dropout layer with a 10\% dropout rate. The output layer contains one sigmoid node to convert input values within 0 to 1. This is the merged probability predicted by the model. We have used the `adam' optimizer and `binary\_crossentropy' loss. The number of epochs was set to 10 (increasing or decreasing epoch more reduced test performance) during model training.

\nd\bf{\ul{Parameter-tuning.}} We have also used grid search to hyper-tune the parameters for each model and presented their best performance. We tuned n\_estimators and max\_depth for RandomForest and ExtraTrees classifier, n\_estimators, and learning\_rate for GradientBoosting and LightGBM, max\_iter for LogisticRegression. For the DNN, we tried varying the number of layers, dropout rate, optimizers, loss (those available in Keras tool for binary classification), and the number of nodes.


The best-performing one was chosen based on the AUC score, as it is mentioned \cite{fan2018early} as the most important evaluation metrics to prioritize code changes for reviewers. The chosen model is then used to compare our longitudinal cross-validation performance with the state-of-the-art. The same classifier is later used to answer our other research questions. The hyper-tuning results for the chosen model are discussed in Section \ref{sec:predcr-hypertuning}.

\nd\bf{\ul{Reproducing state-of-the-art baseline.}} To compare our work with the state-of-the-art \cite{fan2018early}, we have followed the steps presented in their work and also publicly shared git repository \footnote{https://github.com/YuanruiZJU/EarlyPredictionReview} to reproduce it. We have preprocessed our dataset using the same steps as them. Then calculated their features. However, we found a bug in their feature calculations, which calculates the status of some code changes which have not been closed yet. The bug found in author\_feature.py - class AuthorPersonalFeatures - def extract\_features, separates code changes by the creation date of the current code change(for which they are calculating features). However, some of those code changes are merged after the current change was created. When calculating merge ratios, these changes were not excluded. Thus merge ratios now may contain information about the future of those code changes, potentially leaking the target label(merged or abandoned). We have fixed this issue by excluding those open changes when calculating any kind of merge ratio (merged\_ratio, recent\_merged\_ratio, subsystem\_merged\_ratio). We have shared both implementations (original and fixed) in our shared repository \footnote{https://github.com/khairulislam/Predict-Code-Changes.}.

We have used a RandomForest classifier with class\_weight='balanced' as their model, it is equivalent to their usage of RandomForest from the Weka tool with $\alpha$ = 1 in cost ratio. We have hyper-tuned their model and found the best results when n\_estimators are 500 and max\_depth is 5. This hyper-tuned model is then used on their feature set calculated from the same dataset as us, following the longitudinal cross-validation setup. So our results are directly comparable.

\section{Empirical Study Results}
\label{sec:result}
In this section, we answer five research questions:
\begin{enumerate}[label=RQ\arabic{*}., leftmargin=35pt]
  \item Can our proposed approach outperform the state-of-the-art?
  
  \item  How effective is PredCR when only one feature dimension is used?
  
  \item How well does the model handle bias against new authors?
  
  \item How well does our approach work while using multiple revisions?
  
  \item How well does the model improve over time?
  
\end{enumerate}

Answering RQ1 will show how PredCR performs compared to the state-of-the-art
works. It will also validate PredCR's effectiveness in the early prediction of merged
code changes. RQ2 will highlight the performance of each feature dimension used in PredCR. This research question will also validate whether PredCR benefits from using all features rather than using a subset of them. RQ3 will explore whether PredCR has any bias against new authors. We have used author experience-related features, so this may introduce bias against new authors. By answering this research question we have explained how we have handled it and what impact it had on the performance of PredCR. RQ4
is intended to explore if PredCR is able to improve its prediction ability with subsequent revisions. And finally, RQ5 investigates if  PredCR has any advantage of using a longer period of time. With time, training data can be enriched as new code changes are available in a project and hence performance improvement is expected.

\addtocounter{d}{1}
\subsection{Can our proposed approach outperform the state-of-the-art? (\it{RQ\arabic{d})}}
\label{sec:crossValidation}

\subsubsection{Motivation} 
To validate the performance of PredCR, we plan to compare our approach and
performance with the state-of-the-art. The work of Fan et al.
\cite{fan2018early} is considered state-of-the-art on the early prediction of
merged code changes. Their model outperforms the previous works (Jeong et al.
\cite{jeong2009improving}, Gousios et al. \cite{gousios2014exploratory}, etc.)
with respect to most of the metrics. So if PredCR is able
to outperform their models in a similar setup, its superiority and
applicability will be established.

\subsubsection{Approach}

The approach described in Section \ref{sec:experimentationSetup} is used here to evaluate this research question.

\subsubsection{Results}
Table \ref{table:classifierSelection} shows the results for selecting the
best classifier. AUC is chosen as it is suggested to be the best metric for this
task \cite{fan2018early}. The best results for each project are in bold. 
\bf{ We have found that LightGBM has the best overall performance across the three projects.} Which isn't surprising as LightGBM has previously shown better performance than similar gradient boosting decision trees \cite{ke2017lightgbm}. LightGBM showed an AUC score of 86.0, 84.3, and 84.6 for the three projects LibreOffice, Eclipse, and GerritHub, respectively. \bf{We, therefore, have picked LightGBM as the underlying model in our PredCR tool and used it throughout the remaining parts of the paper.}

\begin{table}[t]
  \centering
  \caption{Performance for different classifiers across the three projects}
    \begin{tabular}{lrrr}\toprule
    \multicolumn{1}{l}{\multirow{2}[0]{*}{\textbf{Model}}} & \multicolumn{3}{c}{\textbf{AUC}} \\
    \cmidrule{2-4}
          & \multicolumn{1}{l}{\textbf{LibreOffice}} & \multicolumn{1}{l}{\textbf{Eclipse}} & \multicolumn{1}{l}{\textbf{GerritHub}} \\
          \midrule
    LightGBM & \bf{86.0}  & \bf{84.3}  & \bf{84.6} \\
    DNN   & 85.3  & 82.9  & 84.0 \\
    Random Forest & 85.2  & 83.6  & 83.5 \\
    GradientBoosting (GBT) & 85.4  & 82.8  & 82.7 \\
    ExtraTrees & 85.6  & 83.3  & 83.6 \\
    Logistic Regression & 81.2  & 77.0  & 79.3 \\
    \bottomrule
    \end{tabular}%
  \label{table:classifierSelection}%
\end{table}%

Table \ref{table:crossValidation}, shows the results of our best model and its
comparison with the state-of-the-art \cite{fan2018early}. \textbf{PredCR
outperforms the state-of-the-art in all cases. Our average AUC is around 85\%, where the state-of-the-art is around
71.4\%.} We note that we find a slightly lower performance for Fan et al.~\cite{fan2018early} model compared to its performance as reported in the Fan et al. paper. This can be due to dataset differences and adding a fix in their feature calculation.

\begin{table*}[!htb] \centering
\caption{Longitudinal cross-validation test results with comparison}
\label{table:crossValidation}
\resizebox{\textwidth}{!}{
\begin{tabular}{llrrrrr|rrr}
\toprule
\textbf{Project}  & \textbf{Approach} & \textbf{AUC} & \textbf{ER@20\%} & \multicolumn{3}{c}{\textbf{Merged}} & \multicolumn{3}{c}{\textbf{Abandoned}} \\ \cmidrule{5-10}
& & & &  \textbf{F1(M)} & \textbf{P(M)} & \textbf{R(M)} & \textbf{F1(A)} & \textbf{P(A)} & \textbf{R(A)} \\ 
\cmidrule{1-10}
\multirow{3}{*}{LibreOffice} & Ours & \textbf{86.0} & \textbf{99.3} & \textbf{94.1} & \textbf{95.9} & \textbf{92.4 }& \textbf{48.3} & \textbf{42.2} & \textbf{58.7} \\ \cmidrule{2-10}
& Fan et al \cite{fan2018early} & 70.2 & 96.3 & 86.7 & 91.1 & 82.8 & 31.7 & 26.4 & 42.2 \\
\midrule
\multirow{3}{*}{Eclipse} & Ours &  \textbf{84.3} & \textbf{97.5} & \textbf{92.3} & \textbf{92.9} & \textbf{91.8} & \textbf{57.7} &\textbf{ 56.0} & \textbf{60.0} \\ \cmidrule{2-10}
& Fan et al \cite{fan2018early} & 69.7 & 93.9 & 81.6 & 89.2 & 75.6 & 36.1 & 29.2 & 50.0 \\ 
\midrule
\multirow{3}{*}{GerritHub} & Ours &  \textbf{84.6} & \textbf{99.0} & \textbf{92.0} & \textbf{95.3} & \textbf{88.9} & \textbf{50.0} & \textbf{42.2} & \textbf{62.5} \\ \cmidrule{2-10}
& Fan et al \cite{fan2018early} & 74.4 & 98.2 & 82.9 & 93.9 & 74.8 & 33.4 & 24.1 & 59.0 \\ 
\bottomrule
\end{tabular}
}
\end{table*}

As defined in \sec\ref{sec:evaluationMetrics}, 
we calculate the relative improvement (RIMPR) 
and normalized improvement (NIMPR) \cite{costa2016tipmerge, fan2018early} of PredCR 
over the state-of-the-art baseline (i.e., Fan et al. \cite{fan2018early}). We 
present the improvement of PredCR over the baseline 
in Table \ref{table:crossValidationImprovement} using four metrics AUC, ER@20\%, F1(M), and F1(A). We find that
\begin{inparaenum}[(1)]
    \item PredCR improves the AUC scores by around 14-23\% compared to state-of-the-art \cite{fan2018early} and the normalized improvements are around 46-58\%.
    \item The ER@20\% in state-of-the-art was already around 96\% on average. However, PredCR still provides around 44-81\% normalized improvement.
    \item In terms of f1\_score for merged changes, PredCR provides around 9-13\% relative improvements and 42-58\% normalized improvements. Though the state-of-the-art \cite{fan2018early} were already significant for merged code changes.
    \item For abandoned changes, PredCR improves f1\_score by a large margin. It gives around 50-60\% relative improvements and 24-34\% normalized improvements. Considering only 12\% of the code changes are abandoned, difficulties in accurately predicting the fate of abandoned code changes are significantly higher.
\end{inparaenum}

\begin{table*}[!htb] \centering
\caption{Improvements (RIMPR and NIMPR) of PredCR over the baseline (Fan et al. \cite{fan2018early}) }
\label{table:crossValidationImprovement}
\begin{tabular}{llrr}
\toprule
\textbf{Project} & \textbf{Metric}  & \textbf{RIMPR} & \textbf{NIMPR}  \\ 
\cmidrule{1-4}
\multirow{4}{*}{LibreOffice} & AUC & 22.8 & 53.7  \\ \cmidrule{2-4}
& ER@20\%  & 3.11 & 81.1\\ \cmidrule{2-4}
& F1(M) & 8.53 & 55.6 \\ \cmidrule{2-4}
& F1(A) & 52.4 & 24.3  \\ \midrule
\multirow{4}{*}{Eclipse} & AUC & 20.9 & 48.2 \\ \cmidrule{2-4}
& ER@20\%  & 5.06 & 59.0 \\ \cmidrule{2-4}
& F1(M) & 13.1 & 58.2  \\ \cmidrule{2-4}
& F1(A) & 59.8 & 33.8  \\ \midrule
\multirow{4}{*}{GerritHub} & AUC & 13.7 & 39.8 \\ \cmidrule{2-4}
& ER@20\%  & 0.81 & 44.4\\\cmidrule{2-4}
& F1(M) & 8.56 & 41.5 \\\cmidrule{2-4}
& F1(A) & 49.7 & 24.9 \\ 
\bottomrule
\end{tabular}
\end{table*}

Table \ref{table:featureImportance} shows the importance of our studied features
while running the longitudinal cross-validation process. It was calculated using the
feature\_importances\_ attribute provided by the LightGBM classifier for each
project during the longitudinal cross-validation process and averaged over all runs. The
importance of the top three features for each project is in bold. The review and
project dimensions added in PredCR were not used by Fan et al.
\cite{fan2018early}, but Table \ref{table:featureImportance} shows that they
have a significant impact on prediction performance.

\begin{table*}[!htb]
\centering
\caption{List of features with importance in PredCR }
\label{table:featureImportance}
\resizebox{\textwidth}{!}{
\begin{tabular}{p{1.6cm}lrrr} \toprule
\textbf{Dimension} & \textbf{Feature Name} & \multicolumn{3}{c}{\textbf{Feature Importance}} \\ \cmidrule{3-5}
& & \textbf{LibreOffice} & \textbf{Eclipse} & \textbf{GerritHub} \\ 
\midrule
\multirow{4}{*}{Reviewer} & avg\_reviewer\_experience & \textbf{ 9.67} & 6.73 & \textbf{10.0} \\ 
&  avg\_reviewer\_review\_count \cite{thongtanunam2017review} \cite{baysal2016investigating}  & \textbf{10.9} & \textbf{8.18} & \textbf{8.98} \\ 
&  num\_of\_reviewers \cite{thongtanunam2017review} \cite{jiang2013will}  & 3.64 & 4.94 & 7.73 \\ 
& num\_of\_bot\_reviewers & 2.84 & 0.60 & 1.11 \\
\midrule
\multirow{6}{*}{Author} & author\_merge\_ratio \cite{fan2018early} &  4.72 & 2.68 & 4.24\\ 
&  author\_experience &  \textbf{9.25} & 8.07 & \textbf{8.30} \\ 
&  author\_merge\_ratio\_in\_project \cite{fan2018early} & 1.71 & 3.77 & 1.53 \\ 
&  total\_change\_number \cite{fan2018early} \cite{baysal2016investigating} & 7.23  & \textbf{8.40} & 7.21 \\ 
&  author\_review\_number \cite{fan2018early} \cite{jiang2013will} & 7.61 & \textbf{8.55} & 7.87 \\ 
&  author\_changes\_per\_week \cite{thongtanunam2017review} & 4.91  & 5.39 & 7.02 \\ 
\midrule
\multirow{3}{*}{Project} & project\_changes\_per\_week \cite{thongtanunam2017review} & 7.55 & 7.00 & 7.00\\ 
&  changes\_per\_author  & 5.43 & 6.28 & 4.94 \\ 
& project\_merge\_ratio  & 2.93 & 4.43 & 5.35 \\ 
\midrule
\multirow{4}{=}{Text} & description\_length \cite{fan2018early} & 3.53 & 3.79 & 2.64 \\ 
& is\_bug\_fixing \cite{thongtanunam2017review} \cite{fan2018early} & 0.30 & 0.33 & 0.17\\ 
&  is\_feature \cite{thongtanunam2017review} \cite{fan2018early} & 0.38 & 0.84 & 1.19 \\ 
&  is\_documentation \cite{thongtanunam2017review} \cite{fan2018early} & 0.18 & 0.28 & 0.24\\ 
\midrule
\multirow{8}{*}{Code} & modified\_directories \cite{thongtanunam2017review} & 2.07 & 0.98 & 0.96 \\ 
& subsystem\_num \cite{kamei2013large},  & 2.83 & 5.90 & 3.39 \\
&  modify\_entropy \cite{kamei2013large} & 2.47 & 2.02 & 2.42 \\
& lines\_added \cite{jeong2009improving,fan2018early} & 4.44 & 5.24 & 3.93 \\ 
& lines\_deleted \cite{jeong2009improving, fan2018early} &3.33 & 3.32 & 3.21 \\ 
& files\_modified \cite{gousios2014exploratory} & 1.24 & 1.24 & 1.37  \\ 
&  files\_added \cite{weissgerber2008small, fan2018early}  &  0.48 & 0.80 & 0.72 \\ 
&  files\_deleted \cite{weissgerber2008small,fan2018early} & 0.30 & 0.25 & 0.06 \\ 
\bottomrule
\end{tabular}
}
\end{table*}


\begin{tcolorbox}[flushleft upper,boxrule=1pt,arc=0pt,left=0pt,right=0pt,top=0pt,bottom=0pt,colback=white,after=\ignorespacesafterend\par\noindent]
\nd\it{\bf{RQ1. Can our proposed approach PredCR outperform the state-of-the-art baseline?}} Our PredCR tool is based on the LightGBM model, which 
\textbf{on average}, outperforms the state-of-the-art ~\cite{fan2018early} by 19\% in terms of AUC score. 
If we compare the normalized improvement (NIMPR) metric, PredCR outperforms the state-of-the-art ~\cite{fan2018early} by 48\% in terms of AUC score.  
PredCR outperforms by 10\% for merged and by 54\% for abandoned changes (in terms of F1-score). 
The most informative two features in PredCR are avg\_reviewer\_experience and avg\_reviewer\_review\_count which belong to the Reviewer dimension, none of which were used by Fan et al.~\cite{fan2018early}.  
\end{tcolorbox}

\addtocounter{d}{1}
\subsection{How effective is PredCR when only one feature dimension is used? (\it{RQ\arabic{d})}}
\label{sec:featureDimension}
\subsubsection{Motivation} 
We have described the features we have used in Section
\ref{sec:studiedFeatures}. In this research question, we have investigated how
much performance each feature dimension used in PredCR achieves alone. This will
also validate whether PredCR benefits from using all those feature dimensions or
not.

\subsubsection{Approach}

We have used the same longitudinal ten-fold cross-validation on all projects. We have worked first with all dimensions and later trained and tested the classifier for one feature dimension
only. Then reported the performance metrics.

\subsubsection{Results} 
Table \ref{table:featureDimension} shows PredCR performance using all feature
dimensions and single feature dimension. The best results for each dimension are in bold. The
average AUC on models trained on all dimensions, reviewer, author, project,
text, and code dimensions are 85\%, 77\%, 67\%, 58\%, 53\%, 57\%. In terms of the AUC score, \textbf{PredCR on average improves reviewer, author, project, text, and code models by 10\%, 27\%, 46\%, 60\%, and 49\% respectively.} Except for the reviewer dimension, all other dimensions have poor performance for abandoned code changes.

\begin{table*}[!htb]
\centering
\caption{Performance of PredCR for all features and in each feature dimension }
\label{table:featureDimension}
\resizebox{\textwidth}{!}{
\begin{tabular}{llrrrr}
\toprule
\textbf{Project} & \textbf{Dimension} & \textbf{AUC} & \textbf{ER@20\%} & \textbf{F1(M)} & \textbf{F1(A)} \\ 
\midrule
\multirow{6}{*}{LibreOffice} & All dimensions & \textbf{86.0} & \textbf{99.3} & \textbf{94.1} & \textbf{48.1}  \\ 
& Reviewer & 81.3 & 97.9 & 92.2 & 42.9 \\ 
& Author & 67.7 & 96.1 & 90.7 & 25.1 \\ 
& Project & 50.8 & 91.2 & 76.8 & 11.5 \\ 
& Text & 52.5 & 92.6 & 73.2 & 14.9 \\ 
& Code & 53.7 & 92.6 & 81.0 & 14.5 \\ 

\midrule
\multirow{6}{*}{Eclipse} & All dimensions &  \textbf{84.3} & \textbf{97.5} & \textbf{92.3}  & \textbf{57.7}  \\ 
& Reviewer & 75.9 & 93.3 & 91.5 & 54.2  \\ 
& Author & 65.3 & 92.4 & 81.6 & 31.5 \\ 
& Project & 58.1 & 90.0 & 78.5  & 25.5 \\ 
& Text & 55.1 & 87.7 & 74.0 & 24.2   \\ 
& Code & 55.9 & 88.4 & 76.1 & 24.6  \\ 

\midrule
\multirow{6}{*}{GerritHub} & All dimensions & \textbf{84.6} & \textbf{99.0} & \textbf{91.7} & \textbf{49.3} \\ 
& Reviewer & 72.7 & 95.3 & 86.8 & 35.1 \\ 
& Author & 69.3 & 97.5 & 83.6 & 26.9 \\ 
& Project & 66.4 & 96.8 & 77.7 & 26.2 \\ 
& Text & 52.4 & 90.2 & 58.2  & 19.2\\ 
& Code & 61.2 & 95.8 & 74.8 & 22.5 \\ 
\bottomrule
\end{tabular}
}
\end{table*}

We have presented two examples to demonstrate the importance of the reviewer
dimension and how it affects the change request outcome. In project LibreOffice,
for {change id 65890}\footnote{\url{https://gerrit.libreoffice.org/c/core/+/65890}},
the author was facing build failures because one of the pipeline tests was
failing. 
The reviewer mentioned that the test failed not because of the author's change. If he would have uploaded a
new revision of this patch, the tests might have run successfully. The author later abandoned this change and created another
{change 66203}\footnote{\url{https://gerrit.libreoffice.org/c/core/+/66203}}  for the same
issue. This was later merged successfully with further help from the reviewer.
Clearly, the reviewer's experience directly influenced the outcome of these
changes. In project GerritHub {change id 745519}\footnote{\url{https://review.opendev.org/\#/c/745519/}}, the reviewer suggested that
the change made by the author was unnecessary 
since there was a better alternative.
The experienced reviewer knew about this method, but the author did not. After
reviewer's suggestion he abandoned the change.


To demonstrate the importance of the author dimension, we show an example from our 
dataset below. In LibreOffice
{change id 4071}\footnote{\url{https://gerrit.libreoffice.org/c/core/+/4071}\label{foot:4071}}  the author
gives a fix for several bugs. 
The reviewer compliments the author for fixing this critical problem. The author is
experienced in this project and had been working for more than 1 year, with a
0.98 merge ratio in this project. At the time of this code change, he was making
around 7 code changes per week and also actively reviewing other code changes.


Similarly, here is an example of the project dimension. In LibreOffice
{change id 4071}\footref{foot:4071}, 
the code change is made for the 'core'
sub-project. At the time this code change was created, this sub-project had
around 103 code changes per week, a merge ratio of 0.87, and on average 8 code
changes per developer. The author was making around 7 code changes per week, so
he was a regular developer on that project. We see his code changes get merged
with minimal review.

And, the following example shows the importance of text dimension. 
In project GerritHub {change id
745519}\footnote{\url{https://review.opendev.org/\#/c/745519/}}, the change description says, "add brctl command for neutron-linuxbridge
image". The number of words would be 6. And following the approach of
Thongtanunam et al. \cite{thongtanunam2017review} this code change will be labeled as a feature.

The next one demonstrates the importance of the source code-related dimension. For example, 
LibreOffice {change id 4071}\footref{foot:4071}
is a medium-size code change. The author made 71
line additions and 4 deletions across 6 files. So this is easier for the
reviewers to inspect. We see it gets merged with minimal review. These examples demonstrate the importance of using PredCR with diverse features.

\begin{tcolorbox}[flushleft upper,boxrule=1pt,arc=0pt,left=0pt,right=0pt,top=0pt,bottom=0pt,colback=white,after=\ignorespacesafterend\par\noindent]
\nd\it{\bf{RQ2. How effective is PredCR when only one feature dimension is used?}} The reviewer dimension has the best average AUC score of 77\% across projects for a single dimension. Also, this dimension has moderate performance on abandoned code changes. Author dimension achieved 67\%  AUC score on average. Project dimension achieved on average 58\% AUC score, but there is a significant difference in score between LibreOffice and GerritHub. Text and code dimensions achieved around 53\% and 57\% AUC scores, so their impact is close. Using all dimensions together improved our AUC scores by 10-60\%. This validates that PredCR benefits from the use of all features, compared to its subset.
\end{tcolorbox}

\addtocounter{d}{1}
\subsection{How well does the model handle bias against new authors? (\it{RQ\arabic{d})}}
\label{sec:newAuthor}
\subsubsection{Motivation} 
Table \ref{table:featureImportance} shows high importance of author-related features on PredCR. For a new author, it is more likely to consider him/her as inexperienced and predict a lower possibility for merging. For example, Fan et al. \cite{fan2018early} faced a considerable bias against new authors. Changes made by new authors were mostly being predicted as abandoned. Hence, they had to propose an adjustment approach. They predicted code changes made by new authors using a model that is only trained on code changes by new authors. They used another model trained on all code changes for experienced contributors. We also need to evaluate how much bias PredCR might have for the new authors.

\subsubsection{Approach} 
We have labeled authors with less than ten code changes as new authors following Fan et al. \cite{fan2018early}. Test dataset in each fold of the longitudinal ten-fold cross-validation only contains new authors. Fan et al.'s \cite{fan2018early} results were reproduced using their adjusted approach as they suggested.

\subsubsection{Results} 
Table \ref{table:newAuthor} shows the comparison of PredCR performance with state-of-the-art  \cite{fan2018early}. Our findings from the table are as below,

\begin{enumerate}
    \item PredCR's average AUC score across all projects is 85\% and for new authors, it is 78.7\%. So the performance drop in PredCR for this case is small, considering new authors have either none or few past records. 
    
    \item In terms of AUC scores PredCR improves over Fan et al.'s \cite{fan2018early} adjusted approach for LibreOffice, Eclipse, and GerritHub projects by 26\%, 31\%, and 21\%. The normalized improvements are 43\%, 47\% and 40\%.
    \item In terms of ER@20\%, PredCR provides 5-17\% relative improvements.
    \item For metrics related to merged code changes, PredCR under-performs in terms of F(M) and R(M). This is because PredCR has less bias against abandoned code changes. 
    \item For metrics related to abandoned code changes, PredCR significantly outperforms in terms of F(A) and R(A). But under-performs in terms of P(A). However, this shows that Fan et al.'s \cite{fan2018early} adjusted approach has a considerable bias towards merged code changes.
\end{enumerate}

\begin{table*}[htbp]
\centering
\caption{Performance on changes created by new authors}
\label{table:newAuthor}
\resizebox{\textwidth}{!}{
\begin{tabular}{llrrrrrrrr}
\toprule
\textbf{Project} & \textbf{Approach} & \textbf{AUC} & \textbf{ER@20\%} & \multicolumn{3}{c}{\textbf{Merged}} & \multicolumn{3}{c}{\textbf{Abandoned}} \\ \cmidrule{5-10}
& & & & F1(M) & P(M) & R(M) & F1(A) & P(A) & R(A) \\ 
\midrule
\multirow{2}{*}{LibreOffice} & Ours & \textbf{78.3} & \textbf{94.4} & 52.9 & \textbf{92.3} & 41.6 & \textbf{ 49.6} & 37.1 & \textbf{85.2} \\ 
& Fan et al. \cite{fan2018early} & 62.1 & 83.2 & \textbf{85.3} & 76.0 & \textbf{97.2} & 13.9 & \textbf{49.7} & 8.31\\ 
\midrule
\multirow{3}{*}{Eclipse} & Ours & \textbf{78.9} & \textbf{89.9} & 71.5 & \textbf{87.2} & 61.2 & \textbf{54.0} & 42.7 & \textbf{ 75.8}  \\ 
& Fan et al.\cite{fan2018early} & 60.6 & 76.6 & \textbf{79.4} & 68.7 & \textbf{98.2} & 9.70 & \textbf{54.3} & 5.38 \\ 
\midrule
\multirow{3}{*}{GerritHub} & Ours & \textbf{78.9} & \textbf{91.1} & 64.6 & \textbf{89.8} & 51.9 & \textbf{49.3} & 36.5 & \textbf{78.9}  \\ 
& Fan et al.\cite{fan2018early} & 65.0 & 86.9 & \textbf{84.9} & 75.6 & \textbf{97.1} & 11.3 & \textbf{41.6 }& 6.81 \\ 
\bottomrule
\end{tabular}
}
\end{table*}

We have concluded that in Fan et al.'s \cite{fan2018early} original approach, the bias to experienced authors was introduced by using many features related to the author's past records. For the new authors, these feature values are mostly zero and thus cause a bias against them increasing the likelihood of predicting them as abandoned. Even the adjusted approach ends up having a bias towards merged code changes. To reduce such bias, we have decided not to use the collaborative dimension which considers the collaborative history between author and reviewers. This could have resulted in a decrease in overall performance. However, our addition of features related to reviewer and project dimensions makes up for that deficiency and also improves the overall model performance. 


\begin{tcolorbox}[flushleft upper,boxrule=1pt,arc=0pt,left=0pt,right=0pt,top=0pt,bottom=0pt,colback=white,after=\ignorespacesafterend\par\noindent]
\nd\it{\bf{RQ3. How well does the model handle bias against new authors?}}
PredCR achieved on average 78.7\% AUC score in the longitudinal cross-validation test for new authors, where the state-of-the-art \cite{fan2018early} achieved around 63\%. PredCR gives a more balanced prediction for both classes, while still maintaining a better AUC score. Also, our model performance for new authors (78.7\% AUC) is not far behind the overall model performance (85\% AUC).
\end{tcolorbox}

\addtocounter{d}{1}
\subsection{How well does our approach work while using multiple revisions? (\it{RQ\arabic{d})}}
\label{sec:multipleRevision}
\subsubsection{Motivation}  
So far we have trained and tested with only the initial submission of code. But
in real life, a code change generally goes through several revisions before
finally getting merged or abandoned. Each revision contains updated files based
on reviews received in the previous revisions. Thus an outcome predicted based
on the first revision might be improper for later revisions. The prediction
model needs to be able to update prediction given a code change when a new
revision is pushed. Besides the initial submission, the stakeholders can still
be significantly benefited if a good prediction is available after early-stage
revisions.

Many of the changes are not ready for review during the initial submission. The
reasons can be: (i) build failure, (ii) pipeline test failure, (iii) work in
progress, (iv) merge conflict, (v) unintentionally included changes, and (vi)
dependent on any other change. For this, the author has to push more patches.
Multiple patches are already uploaded before the review even starts. A merge
prediction made only on the first patch would miss any of these cases. For
example, in project Eclipse, for {change-id 167412}\footnote{\url{https://git.eclipse.org/r/c/platform/eclipse.platform.swt/+/167412}},
the initial patch was labeled
work in progress. 
The second patch
faced a build fail. On the third patch, it was labeled as ready-for-review and
finally was merged after the eighth patch. In project LibreOffice, for {change-id 100373}\footnote{\url{https://gerrit.libreoffice.org/c/core/+/100373}}, it took the
author five patches to fix build fails. Only then the change was ready for review
and finally was merged at the sixth patch.


\subsubsection{Approach} 

We have designed two adjusted approaches for the merge prediction of a code change in revision rounds. In the first approach, we have added only the review number to the existing feature set. This approach does not train on any previous activities within the patchset. In the second approach, we have added features related to reviews and other activities of previous revisions in the feature set. Similar approach was followed by Gousios et al. \cite{gousios2014exploratory}, Jiang et al. \cite{jiang2013will} and Thongtanunam et al. \cite{thongtanunam2017review}. In both approaches, we have used the longitudinal data setup during validation. Change features are sorted according to their creation time. 

We have used two different approaches because they will show how PredCR performs with or without considering review activities from previous revisions.\textbf{ One important difference is that for this approach our features are calculated right after a new revision is uploaded. So that we can give updated predictions on the code change before reviewers have to do any review.} Both Gousios et al. \cite{gousios2014exploratory} and Thongtanunam et al. \cite{thongtanunam2017review} calculated features just before the pull request or the code change is closed.  However reviews are already done at that point, so predicting at that point would not be helpful for reviewers.

\subsubsection{Results} 

Table \ref{table:multipleRevision1} shows the test results with the first approach. Column RIMPR and NIMPR show the improvement and normalized improvement \cite{costa2016tipmerge, fan2018early} in the AUC scores at the last revision compared to the first.  Here we have added  'revision\_number' in the feature set so that the model knows at which stage of review this code change belongs. Jiang et al. \cite{jiang2013will} used patch\_no when predicting whether a patchset will be accepted in the git repository of the Linux kernel. Note that this result is not comparable with the one shown in Table \ref{table:crossValidation} because the test set is different. However, the average AUC is still significant. 

'Total' presents results when the test fold contains all changes of that fold. 'First revision' presents the result when the test fold only contains changes at their first revision. The last revision means when the code change was finally merged or abandoned. 'Last revision' presents the result when the test fold only contains changes at that revision. Table \ref{table:multipleRevision1} shows that in all cases the AUC score has improved in the last revision. That is expected because the fate of the code change is almost set at that time.   

\begin{table}[htbp]
\centering
\caption{AUC(\%) for multiple revisions with revision number (Approach 1)}
\label{table:multipleRevision1}
\resizebox{\textwidth}{!}{
\begin{tabular}{lrrrrrr}
\toprule
\textbf{Project} & \textbf{Total} & \textbf{First revision} & \textbf{Last revision} & \textbf{RIMPR} & \textbf{NIMPR } \\ 
\midrule
LibreOffice & 85.2 & 86.2 & 92.5 & 7.7 & 47 \\
Eclipse & 77.8 & 83.7 & 86.1 & 2.9 & 15 \\
GerritHub & 82.0 & 85.1 & 86.5 & 1.6 &  9.4 \\
\bottomrule
\end{tabular}
}
\end{table}

For the second approach we have used revision number \cite{jiang2013will}, weighted\_approval\_score, avg\_delay\_between\_revisions \cite{thongtanunam2017review}, and number\_of\_messages as extra features. Weighted approval score is calculated at each revision by adding label values of previous revisions multiplying by the fraction of current\_revision\_no and current\_revision\_no + 1. This will add more weight to the labels in the later revisions. \\Avg\_delay\_between\_revisions is calculated in days. Table \ref{table:multipleRevision2} shows the average test AUC scores achieved during the experiments. Column RIMPR and NIMPR show the improvement 
and normalized improvement\cite{costa2016tipmerge, fan2018early} of the AUC scores at the last revision compared to the first. 

\begin{table}[htbp]
\centering
\caption{AUC(\%) for multiple revisions with previous revision related features (Approach 2)}
\label{table:multipleRevision2}
\resizebox{\textwidth}{!}{
\begin{tabular}{lrrrrr}
\toprule
\textbf{Project} & \textbf{Total} & \textbf{First revision} & \textbf{Last revision} & \textbf{RIMPR} & \textbf{NIMPR} \\ 
\midrule
LibreOffice & 88.2 & 86.2 & 98.8 & 15 & 92 \\
Eclipse & 79.5 & 83.7 & 89.0 & 6.1 & 33 \\
GerritHub & 82.6 & 85.1 & 90.0 & 5.8 & 33 \\
\bottomrule
\end{tabular}
}
\end{table}

Overall AUC scores and AUC scores in the last revision both have improved in this approach. During the first revision, these previous revision-related features do not exist. However, this result shows that adding previous revision-related features can improve prediction performance in later revisions. Also we have found, for changes with only one revision the AUC scores are 86\%, 85.2\%, and 83.5\% in project LibreOffice, GerritHub, and Eclipse. But for changes with multiple revisions, their AUC scores at the last revision (when the change was finally closed) are 98.6\%, 89.4\%, and 86\% respectively.
But since our primary goal is to give better results during the initial submission, we have not focused too much on this point.

\begin{tcolorbox}[flushleft upper,boxrule=1pt,arc=0pt,left=0pt,right=0pt,top=0pt,bottom=0pt,colback=white,after=\ignorespacesafterend\par\noindent]
\nd\it{\bf{RQ4. How well does our approach work while using multiple revisions?}} 
PredCR achieves around 78-88\% AUC score when predictions are updated at the submission of each new revision. PredCR can improve prediction at the last revision up to 8\%, compared to the prediction performance at the first revision without using previous revision activity-related features. Using previous revision activity-related features can improve the performance up to 15\%. So PredCR can be adjusted with significant results to update predictions at later revision.
\end{tcolorbox}

\addtocounter{d}{1}
\subsection{ How well does the model improve over time? (\it{RQ\arabic{d})}}
\label{sec:improvementOverTime}
\subsubsection{Motivation}  
In real-life scenarios, the number of changes will keep increasing over time.
Hence, the model can be trained on a larger dataset. But it is important to
validate whether increasing the size of the training dataset will increase the
performance of PredCR.

\subsubsection{Approach} 
We have followed a longitudinal ten-fold cross-validation setup to calculate
model performance in each project. As explained in the approach of RQ1, this
validation setup ensures no future data is used during training. The code
changes are sorted by their time of creation and the model trained on past data
is used to predict future code changes. The performance of subsequent folds of
validation presents the outcome of the model over time. Therefore, we have used
the results achieved in each fold of the longitudinal cross-validation setup performed in
RQ1, to validate whether PredCR performance improves in later folds.

\subsubsection{Results}
Table \ref{table:improvementOverTime} shows the prediction performance of the
model during each fold by AUC score. The results show that the performance does
not monotonically increase over time. However, the performance in the last half
is better on average than that in the first half. In the last fold, both
LibreOffice and GerritHub achieved the best results. Eclipse achieved the best
AUC score in the 6th fold. {Average AUC score for LibreOffice, Eclipse, and GerritHub in fold 1-5 are respectively  82\%, 82\%, and 81.4\%. And in fold
6-10 they are 89.2\%, 86.0\%, and 87.9\%. So AUC
scores on average improved  9\%, 5\%, and 8\% in the latter half of the longitudinal cross-validations.}

\begin{table}[htbp]
\centering
\caption{AUC score in each fold}
\label{table:improvementOverTime}
\begin{tabular}{rrrr}\toprule
\textbf{Fold} & \textbf{LibreOffice} & \textbf{Eclipse} & \textbf{GerritHub} \\ 
\midrule
1 & 82.6 & 76.6  & 86.4  \\ 
2 & 80.7 & 82.1  & 72.9 \\ 
3 & 81.0 & 81.2 & 80.6  \\ 
4 & 80.3 & 86.9 & 79.7 \\ 
5 & 87.0 & 86.1 & 87.5  \\ 
6 & 87.5 & \textbf{88.6} & 84.8 \\ 
7 & 88.2 & 84.3 & 85.5  \\ 
8 & 89.6 & 85.6 & 89.6 \\ 
9 & 90.9 & 87.1 & 88.4 \\ 
10 & \textbf{91.8} & 84.9 & \textbf{91.1}  \\ 
\bottomrule
\end{tabular}
\end{table}

\begin{tcolorbox}[flushleft upper,boxrule=1pt,arc=0pt,left=0pt,right=0pt,top=0pt,bottom=0pt,colback=white,after=\ignorespacesafterend\par\noindent]
\nd\it{\bf{RQ5. How well does the model improve over time?}} 
The longitudinal cross-validation setup sorts data by time and after each fold, one more window is added to the training data, so train data size increases too. In this real-world scenario, PredCR has improved 5-9\% in terms of AUC scores in the latter half of the fold. This validates that, in an active project, with the passage of time, PredCR will be able to improve its performance as more changes are created. 
\end{tcolorbox}

\section{Discussions}\label{sec:discussions}
In this section, we first offer more detailed insights into the performance of PredCR by analyzing the performance 
based on hyper-parameters and run-time (\sec\ref{sec:discuss-predcr-performance}). We then discuss 
the implications of PredCR and our study findings to the field of software engineering practitioners and 
research in \sec\ref{sec:implications}.

\subsection{A Deeper Dive Into PredCR Performance}\label{sec:discuss-predcr-performance}
In \sec\ref{sec:costEffectiveness}, we first analyze the effectiveness of PredCR based on the presence of more/fewer code changes. 
In \sec\ref{sec:timeEfficiency}, we report how much time PredCR takes to train. 
In \sec\ref{sec:predcr-hypertuning}, we report how the performance of PredCR changes based on different values of hyper-parameters. 
We have used the PredCR in \sec\ref{sec:result} after the hyper-parameter tuning. In Section \ref{sec:excludingDimensions} we discuss our model results after excluding each dimension from the feature set. Finally, Section\ref{sec:developerEffort} shows the efforts developers spent per code changes in our dataset.

\subsubsection{Effectiveness of PredCR with gradual increase in code changes}
\label{sec:costEffectiveness}
In this section, we will investigate the performance of PredCR with an increased percentage of inspected code changes. Since reviewing code changes is a costly and time-consuming task, it is not feasible to inspect all the reviews. Like previous studies, we use 20 as the default value for $K$ in ER@K\%. To observe the performance of PredCR with increased $K$, we increase the value from 10 to 90 and repeat the experiment. Table~\ref{table:costEffectiveness} presents that PredCR outperforms Fan et al.~\cite{fan2018early} for all the projects at every $K$ value. Though the ER is supposed to decrease as $K$ increases (the number of abandoned changes increases in the top $K$\% of the list), still PredCR performs well.  

\begin{table}[htbp]
\centering
\caption{ER@20\% for different $K$}
\label{table:costEffectiveness}
\begin{tabular}{rrrrrrr}
\toprule
\textbf{K} &  \multicolumn{2}{c}{\textbf{LibreOffice}} & \multicolumn{2}{c}{\textbf{Eclipse}} & \multicolumn{2}{c}{\textbf{GerritHub}} \\ \cmidrule{2-7}
& \textbf{Ours} & \textbf{Fan et al\cite{fan2018early}} & \textbf{Ours} & \textbf{Fan et al\cite{fan2018early}} & \textbf{Ours} & \textbf{Fan et al\cite{fan2018early}} \\ 
\midrule

10 & \textbf{99.5} & 97.3 & \textbf{97.9} & 95.2 & \textbf{99.4}  & 98.2 \\ 
20 & \textbf{99.2} & 96.3 & \textbf{97.5} & 93.9 & \textbf{99.0} & 98.2 \\ 
30 & \textbf{98.9} & 95.1 & \textbf{96.8} & 93.1 & \textbf{98.6} & 97.3 \\ 
40 & \textbf{98.6} & 94.4 & \textbf{96.4} & 92.4 & \textbf{98.0} & 96.3\\ 
50 & \textbf{98.1} & 93.7 & \textbf{95.8} & 91.4 & \textbf{97.3} & 95.8 \\ 
60 & \textbf{97.8} & 92.8 & \textbf{95.2} & 90.5 & \textbf{96.8} & 94.7 \\ 
70 & \textbf{97.4} & 92.0 & \textbf{94.4} & 89.4 & \textbf{96.2} & 93.8 \\ 
80 & \textbf{96.6} & 90.0 & \textbf{93.3} & 88.1 & \textbf{95.5} & 92.5 \\ 
90 & \textbf{95.7}  & 89.7 & \textbf{91.8} & 85.9 & \textbf{94.3} & 91.2 \\ 
\bottomrule
\end{tabular}
\end{table}


\subsubsection{Time efficiency}
\label{sec:timeEfficiency}

In this section, we discuss the time needed to train the model and its
prediction time. If the model takes too long to predict, then the reviewers
would not get the updated predictions in time, thus discouraging them from applying it.
Moreover, new changes keep coming, and it would be challenging to update the
model if it takes too long. Our used environment provides 16 Gigabytes of Ram, 4 CPU cores, and a 2.20GHz Intel Xeon CPU. In Table
\ref{table:timeEfficiency}, we have presented model training times in seconds. For LibreOffice, Eclipse, and GerritHub,
PredCR training across all 10 folds takes on average 3.56, 2.39, and 2.37
seconds. Where the state-of-the-art~\cite{fan2018early} takes on average 6.97, 14.2, and
6.60 seconds. 

\begin{table}[htbp]
\centering
\caption{Model training time (seconds) in each fold}
\label{table:timeEfficiency}
\begin{tabular}{rrrrrrr}
\toprule
\textbf{Fold} & \multicolumn{2}{c}{\textbf{Libreoffice}} & \multicolumn{2}{c}{\textbf{Eclipse}} & \multicolumn{2}{c}{\textbf{GerritHub}} \\ \cmidrule{2-7}

& \textbf{Ours} & \textbf{Fan's} & \textbf{Ours} & \textbf{Fan's} & \textbf{Ours} & \textbf{Fan's}  \\ \midrule
1 & 1.64 & 1.99 & 1.22 & 2.85 & 1.28 & 1.99  \\ 
2 & 2.13 & 3.09 & 1.65 & 5.29 & 1.77 & 3.01 \\ 
3 &2.69 & 4.20 & 1.82 & 7.68 &  1.79  & 4.07\\ 
4 & 3.07 & 5.24 & 1.98 & 10.1 & 2.06 & 5.04 \\ 
5 & 3.30 & 6.32 & 2.21 & 12.5 & 2.38 & 6.00  \\ 
6 & 3.78 & 7.58 & 2.46 & 15.3 & 2.49 & 7.10 \\ 
7 & 4.19 & 8.65 & 2.71 & 18.0 & 2.87 & 8.18 \\ 
8 & 4.61 & 9.75 & 3.08  & 20.7 & 2.90 & 9.13\\ 
9 & 5.02 & 10.9 & 3.22 & 23.4 & 2.94 & 10.2 \\ 
10 & 5.17 & 12.0 & 3.54 & 25.8 & 3.24 & 11.3 \\ 
Average & \textbf{3.56} & \textbf{6.97}  & \textbf{2.39} & \textbf{14.2} & \textbf{2.37} & \textbf{6.60}\\
\bottomrule
\end{tabular}
\end{table}
\addtocounter{o}{1}


\subsubsection{Impact of Hyper tuning of PredCR}\label{sec:predcr-hypertuning}
Hyper-tuning the parameters of the selected classifier is needed to ensure best model performance during practical use \cite{probst2019tunability}. We have hyper-tuned the selected LightGBM classifier with varying the number of estimators and learning rate. The results are shown in Table \ref{table:hypertuning}. The best parameters are n\_estimators= 500 and learning\_rate= 0.01 based on AUC score.

\begin{table}[!htb] \centering
\caption{Hyper-tuning of PredCR}
\label{table:hypertuning}
\resizebox{\textwidth}{!}{
\begin{tabular}{p{1.6cm}|rrrrr}\toprule
\textbf{Project} & \textbf{n\_estimators}  & \textbf{learning\_rate} & \textbf{AUC}  &
\textbf{F1(M)} & \textbf{F1(A)} \\ 
\midrule
\multirow{4}{=}{LibreOffice} & 100 & 0.1  & 84.8 & 94.9 & 48.5 \\ \cmidrule{2-6}
& 100 & 0.01  & 85.6 & 91.7 & 43.4 \\ \cmidrule{2-6} 
& 500 & 0.1  & 83.3 & \textbf{96.3} & \textbf{48.8} \\ \cmidrule{2-6}
& 500 & 0.01  & \textbf{86.0} & 94.2 & 48.1  \\ 
\midrule
\multirow{4}{=}{Eclipse} & 100 & 0.1 & 84.0  & 92.6 & 57.6 \\ \cmidrule{2-6}
& 100 & 0.01 & 83.4 & 90.9 & 55.2\\ \cmidrule{2-6} 
& 500 & 0.1 & 83.1 & \textbf{93.9} & \textbf{59.4} \\ \cmidrule{2-6} 
& 500 & 0.01 & \textbf{84.3} & 92.3 & 57.7 \\ 
\midrule
\multirow{4}{=}{GerritHub} & 100 & 0.1 & 83.8 & 93.3 & 51.6 \\ \cmidrule{2-6}
& 100 & 0.01 & 83.9 & 89.1 & 44.2 \\ \cmidrule{2-6} 
& 500 & 0.1 & 82.8 & \textbf{95.8 } & \textbf{56.2}  \\\cmidrule{2-6} 
& 500 & 0.01 & \textbf{84.6} & 92.0 & 50.0 \\

\bottomrule
\end{tabular}
}
\end{table}

\subsubsection{Impact of excluding each dimension \label{sec:excludingDimensions}}

This section presents the importance of each feature dimension. We exclude one feature at a time and rerun the experiment.
From Table \ref{table:excludingFeatureDimension} we can see that both reviewer and author dimensions have significant impacts on the model performance. However, removing the author dimension still gives around 82\% AUC score. So if potential bias against new authors becomes a concern, removing this dimension will not make the model unusable.

\begin{table*}[!htb]
\centering
\caption{Performance of PredCR after excluding one feature dimension at a time}
\label{table:excludingFeatureDimension}
\resizebox{\textwidth}{!}{
\begin{tabular}{llrrrr}
\toprule
\textbf{Project} & \textbf{Excluded Dimension} & \textbf{AUC} & \textbf{ER@20\%} & \textbf{F1(M)} & \textbf{F1(A)} \\ 
\midrule
\multirow{5}{*}{LibreOffice} & Reviewer & 70.0 & 96.9 & 91.7 & 26.8 \\ 
& Author & 82.4 & 98.4 & 93.5 & 45.7  \\ 
& Project & 85.9 & 99.2 & 93.8 & 46.8 \\ 
& Text & 85.9 & 99.2 & 94.0 & 47.9 \\ 
& Code & 85.7 & 99.1 & 93.9 & 47.6 \\ 

\midrule
\multirow{5}{*}{Eclipse} & Reviewer & 68.4 & 94.1 & 84.1 & 33.4  \\ 
& Author & 81.1 & 96.4 & 92.9 & 56.4 \\ 
& Project & 84.0 & 97.3 & 91.9  & 57.3 \\ 
& Text & 84.1 & 97.5 & 92.2 & 57.3  \\ 
& Code & 83.6 & 97.1 & 92.3 & 57.1  \\ 

\midrule
\multirow{6}{*}{GerritHub} & Reviewer & 73.3 & 97.9 & 84.6 & 30.7 \\ 
& Author & 82.8 & 98.5 & 91.3 & 47.8 \\ 
& Project & 83.5 & 98.9 & 91.6 & 47.7 \\ 
& Text & 84.4 & 98.9 & 91.8 & 49.4 \\ 
& Code & 84.5 & 98.9 & 91.8 & 49.5 \\ 
\bottomrule
\end{tabular}
}
\end{table*}

\subsubsection{Developer effort spent for changes \label{sec:developerEffort}}
In this section, we show how much effort the developers on average spent on code changes. We consider its duration in days, the number of messages, and the number of revisions as effort.  Duration is measured as the number of days spent from the creation of the code change till it gets merged or abandoned. We found there were occasional large values of these metrics and removing such outliers as noises is a standard statistical process \cite{tukey1977exploratory}. Following Tukey et al. \cite{tukey1977exploratory}, we have removed values outside these two ranges: (a) Lower limit: first quartile - 1.5 * IQR (b) Upper limit: third quartile + 1.5 * IQR. Where IQR (Inter Quartile Range) is calculated by subtracting the first quartile from the third. After removing the outliers, we calculated the mean of those values and presented them in Table \ref{table:developerEffort}. From Table \ref{table:developerEffort} we see, abandoned changes are generally taking more time to close, have fewer messages and revisions per change. These stats are consistent with the results found by Wang et al. \cite{wang2019my} who investigated in detail why code changes get abandoned.

\begin{table*}[htbp]
\centering
\caption{Developer effort spent on the code changes}
\label{table:developerEffort}
\resizebox{\textwidth}{!}{
\begin{tabular}{llrrr}
\toprule
\textbf{Project} & \textbf{Approach} &\textbf{Duration in days} & \textbf{Messages} & \textbf{Revisions} \\  \midrule

\multirow{3}{*}{LibreOffice} & Total & 0.90 & 5.81 & 2.28 \\ 
& Merged & 0.89 & 5.87 & 2.36 \\ 
& Abandoned & 0.98 & 5.21 & 1.42 \\
\midrule
\multirow{3}{*}{Eclipse} & Total & 2.11 & 8.75 & 2.24  \\ 
& Merged & 2.09 & 9.18 & 2.36 \\ 
& Abandoned & 2.30 & 6.43 & 1.62  \\
\midrule
\multirow{3}{*}{GerritHub} & Total & 1.60 & 9.16 & 2.00  \\ 
& Merged & 1.57 & 9.37 & 2.04 \\ 
& Abandoned & 1.90 & 7.50 & 1.68 \\
\bottomrule
\end{tabular}
}
\end{table*}

\subsubsection{Cross Project Performance \label{sec:crossProject}}
For new projects, there might not be enough data to start giving predictions. In those cases, models pre-trained on other projects might be useful during the initial stage of the project. Here we have evaluated PredCR performance in cross-project settings. We have trained the model on a complete dataset of one project and tested it on a complete dataset from another project.

\begin{table*}[!htb]
\centering
\caption{Performance of PredCR across projects }
\label{table:acrossProject}
\resizebox{\textwidth}{!}{
\begin{tabular}{llrrrr}
\toprule
\textbf{Source Project} & \textbf{Target Project} & \textbf{AUC} & \textbf{ER@20\%} & \textbf{F1(M)} & \textbf{F1(A)} \\ 
\midrule
\multirow{2}{*}{LibreOffice} & Eclipse & 64.3 & 95.1 & 85.2 & 23.0 \\ 
& GerritHub & 66.9 & 92.5 & 88.7 & 32.2 \\ \midrule

\multirow{2}{*}{Eclipse} & LibreOffice & 77.5 & 97.5 & 84.6 & 30.6  \\ 
& GerritHub & 76.6 & 97.8 & 88.0 & 36.0  \\ \midrule

\multirow{2}{*}{GerritHub} & LibreOffice & 79.2 & 98.3 & 84.4 & 30.7  \\ 
& Eclipse & 81.1 & 95.9 & 87.6 & 58.2 \\ 
\bottomrule
\end{tabular}
}
\end{table*}

From Table \ref{table:acrossProject} we see PredCR maintains around considerable performance even across projects. So PredCR pre-trained on other projects can be effectively used for new projects. Notice that this result is not comparable to the ones from Section \ref{sec:result} as it doesn't follow longitudinal cross-validation.

\subsection{Implications of Findings}
\label{sec:implications}



As described in \sec\ref{sec:introduction},
the basic usage scenario of our tool is to provide early warnings to authors, reviewers, and the management about
review iterations that will eventually be abandoned. As such, PredCR can be effective 
for the diverse major stakeholders in software engineering: 
\begin{inparaenum}[(1)]
\item \bf{Project Manager and leads} to prioritize code review and code change 
efforts based on the recommendation from PredCR,  
\item \bf{Software Developers} to save time and efforts by focusing on code changes that will most likely be merged (as predicted by PredCR), and 
\item \bf{Software Engineering Researchers} to further investigate useful features like reviewer dimension in relevant early prediction tools.  
\end{inparaenum}


\bf{\ul{Project Manager.}} This tool can provide benefits to software project
management. If the management can predict the negative outcome of a
review-iteration early, they may analyze the cause and take necessary steps if
required. Multiple reasons may act behind the abandonment of changes such as
resource allocation, job environment, efficiency mismatch between the author and
the reviewer, and even their relations. Some of these may be addressed well by
the early intervention of the management and thus revert the result of a
particular review-iteration. Thus the company may save lots of time and
resources. 

Indeed, code review is a very important aspect of modern software engineering. Large
software companies, as well as Open Source projects, are practicing it with care.
Researchers are trying to generate insight from large repositories of code
review and try to bring efficiency in the process to save the cost of production. In
this work, we study a relatively under-studied problem of predicting whether a
code change will eventually be merged or abandoned at an early stage of the code
review process. We design a machine learning model to apply carefully selected
features generated as a result of communication between the developers and the
reviewers. Our developed tool PredCR is expected to save wastage of effort or
help to recover from being abandoned by inviting early intervention of the
management.

\bf{\ul{Software Developers.}} The code review process requires serious effort
and also is time-consuming. The authors and reviewers involved in a review
iteration are likely to get frustrated if they see that their effort goes in
vain, i.e., a patchset has to be abandoned wasting their efforts for quite some
time. If they get an early indication from our tool that their current review
process is going in a negative direction, they may become cautious, seek
external/management help, or at least be prepared mentally. If the management
makes a decision early, their efforts would be saved. Thus it would benefit the
practitioners.


\bf{\ul{Software Engineering Researchers.}} Prior SE researchers followed different approaches including statistical
methods, parametric models, and machine learning (ML) methods for software
effort and duration prediction \cite{shepperd2001using}, software cost
prediction \cite{jorgensen2007systematic}, software fault or defect prediction
\cite{hall2012systematic,shepperd2014researcher}, etc. Search-based peer
reviewer recommendation \cite{ouni2016search} is another related area. Different
prediction models were introduced in the SE (Software Engineering) domain such as predicting the question
quality \cite{baltadzhieva2015predicting} and response time
\cite{goderie2015eta} in Stack Overflow. In these models, the authors exploit
the interactions among users in different contexts related to software
engineering. Bosu et al. \cite{bosu2015characteristics} identified factors that lead to
useful code reviews. Some prior research suggests that a higher number of
reviewers reduces the number of bugs and increases the probability of
acceptability \cite{bavota2015four,kononenko2015investigating}. The experience of
the reviewers sometimes leads to useful code changes
\cite{kononenko2015investigating,baysal2013influence}.


\section{Threats to Validity}
\label{sec:threatsToValidity}

\textbf{\ul{Threats to internal validity}} refers to errors in our implementation. We have cross-checked all data mined to ensure the data used is valid and contains all changes available within that period. We have also removed changes for which full data was not available (i.e. some old changes were missing patch data from Gerrit response). We have rerun the pre-processing steps several times to ensure the same statistics of the final dataset. We have removed changes for which the outcome is obvious (i.e. changes labeled ``WIP" or ``DO NOT MERGE" etc). So that the dataset only contains changes for which prediction is needed. To make the comparisons compatible with the state-of-the-art, we have followed the process presented in their work and reproduced their feature set and experimentation. Our experiments follow the longitudinal setup, which prevents past data to be used in training. This same setup has been followed by prior works \cite{fan2018early, jeong2009improving} in similar scenarios. As the dataset is an imbalanced one, we need to prevent the model from being biased on the majority class. We have balanced classification loss to counter the data imbalance problem. We have also shown the effect of hyper-tuning on model performance. Then made comparisons with the state-of-the-art \cite{fan2018early} using all metrics presented by them. 

\textbf{\ul{Threats to external validity}} refers to the generalization of our tool. For PredCR, it is validated by our test results for unseen data (Section \ref{sec:crossValidation}).  Also despite having large feature importance for experience-related features, for new contributors, PredCR still outperforms the state-of-the-art (Section \ref{sec:newAuthor}). So there is no threat to use this model even developers who are new or past tracks are missing.  Also with the increase of the training dataset, we have shown a positive impact in test results in later half folds of our longitudinal cross-validation result ( Section \ref{sec:improvementOverTime}). Even when prediction is updated for each revision, PredCR shows significant performance (Section \ref{sec:multipleRevision}). With different numbers of $K$, PredCR will still show better predictions than the state-of-the-art (Section \ref{sec:costEffectiveness}). PredCR takes little time to train and test (Section \ref{sec:timeEfficiency}) which validates its practical usability. We have also shown in Section \ref{sec:crossProject} PredCR pre-trained on a project can still perform well for external projects. Even within projects, we found there are sub-projects from different domains and new sub-projects keep getting added to the project over time. PredCR still maintains a significant overall performance. However, the prioritization given by PredCR doesn't imply the importance of the code change or how fast it will be closed. So users need to be cautious when using PredCR in such scenarios.

\textbf{\ul{Threats to construct validity}} refers to the suitability of our evaluation metrics. We have used the evaluation metrics following prior works in the same domain. Both AUC and cost-effectiveness have been widely used in the prediction models of software engineering studies \cite{fan2018early, jiang2013personalized, xia2015elblocker}. We have presented precision, recall, and f1-scores for both classes so that model performance for both of them can be investigated. Also, the metrics have been calculated after averaging over multiple runs of the experimentation. So we believe there is little threat to the validity of our work in practice.    

\section{Conclusion}
\label{sec:conclusion}
Modern code review is an integral part to ensure the quality and timely delivery of software systems. 
Unfortunately, around 12\% of the code changes in a software system are abandoned, i.e, they are not merged to the main code base of the software system. As such, any tool to detect such abandoned changes well in advance can assist software teams with reduced time and effort (e.g., by prioritizing code changes for review that is most likely going to be merged).  In this paper, we present a tool named PredCR that can predict at an early stage of a code review iteration whether a code change would be merged or abandoned
eventually.  This tool is developed using a LightGBM-based classifier following
a supervised learning approach including features related to the reviewer,
author, project, text, and code changes and a dataset of 146,612 code changes
from three Gerrit open source projects. PredCR outperforms the state-of-the-art tool by Fan et al.~\cite{fan2018early} by 14-23\% (in terms of AUC score) and
achieves around 85\% AUC score on average. We have conducted an empirical study on
the applicability of PredCR. We find that the new features like reviewer dimensions that are introduced in PredCR are the most informative. We also find that compared to the baseline, PredCR is more 
effective towards reducing bias against new developers. PredCR uses historical data in the code review repository and as such the performance of PredCR improves as a software system evolves with new and more data. Therefore, PredCR offers more accuracy over the state-of-the-art baseline to early 
predict merged/abandoned code changes in diverse use cases. As such, PredCR can help to reduce the waste of time and efforts
of all stakeholders (e.g., program author, reviewer, project management, etc.)
involved in code reviews with early prediction, which can be used to prioritize efforts and time during the triaging of code changes for reviews.



\begin{small} 
\bibliographystyle{elsarticle-num}
\bibliography{bibliography}   
\end{small}
%
%
%
%

\end{document}